\documentclass[twocolumn]{aastex63}

\usepackage{mathtools}
\usepackage{amsmath}
\received{}
\revised{}
\accepted{}
\submitjournal{ApJ}

\shorttitle{Impact of rotation and compactness on black holes}
\shortauthors{Mapelli et al.}

\begin{document}

\title{Impact of the Rotation and Compactness of Progenitors on the Mass of Black Holes}

\correspondingauthor{Michela Mapelli}
\email{michela.mapelli@unipd.it}

\author[0000-0001-8799-2548]{Michela Mapelli}
\affiliation{Physics and Astronomy Department Galileo Galilei, University of Padova, Vicolo dell'Osservatorio 3, I--35122, Padova, Italy}
\affiliation{INFN-Padova, Via Marzolo 8, I--35131 Padova, Italy}
\affiliation{INAF-Osservatorio Astronomico di Padova, Vicolo dell'Osservatorio 5, I--35122, Padova, Italy}

\author[0000-0003-0930-6930]{Mario Spera}
\affiliation{Physics and Astronomy Department Galileo Galilei, University of Padova, Vicolo dell'Osservatorio 3, I--35122, Padova, Italy}
\affiliation{Center for Interdisciplinary Exploration and Research in Astrophysics (CIERA), Evanston, IL 60208, USA}
\affiliation{Department of Physics \& Astronomy, Northwestern University, Evanston, IL 60208, USA}
\affiliation{INFN-Padova, Via Marzolo 8, I--35131 Padova, Italy}

\author{Enrico Montanari}
\affiliation{Physics and Astronomy Department Galileo Galilei, University of Padova, Vicolo dell'Osservatorio 3, I--35122, Padova, Italy}

\author[0000-0003-0636-7834]{Marco Limongi}
\affiliation{Istituto Nazionale di Astrofisica - Osservatorio Astronomico di Roma, Via Frascati 33, I-00040, Monteporzio Catone, Italy} 
\affiliation{Kavli Institute for the Physics and Mathematics of the Universe, Todai Institutes for Advanced Study, the University of Tokyo, Kashiwa, Japan 277-8583 (Kavli IPMU, WPI)}   

\author[0000-0002-3589-3203]{Alessandro Chieffi}
\affiliation{Istituto Nazionale di Astrofisica - Istituto di Astrofisica e Planetologia Spaziali, Via Fosso del Cavaliere 100, I-00133, Roma, Italy} 
\affiliation{Monash Centre for Astrophysics (MoCA),
School of Mathematical Sciences, Monash University, Victoria 3800, Australia}

\author[0000-0002-8339-0889]{Nicola Giacobbo}
\affiliation{Physics and Astronomy Department Galileo Galilei, University of Padova, Vicolo dell'Osservatorio 3, I--35122, Padova, Italy}
\affiliation{INFN-Padova, Via Marzolo 8, I--35131 Padova, Italy}
\affiliation{INAF-Osservatorio Astronomico di Padova, Vicolo dell'Osservatorio 5, I--35122, Padova, Italy}

\author[0000-0002-7922-8440]{Alessandro Bressan}
\affiliation{SISSA, via Bonomea 265, I--34136 Trieste, Italy}
\affiliation{INAF-Osservatorio Astronomico di Padova, Vicolo dell'Osservatorio 5, I--35122, Padova, Italy}

\author[0000-0003-3462-0366]{Yann Bouffanais}
\affiliation{Physics and Astronomy Department Galileo Galilei, University of Padova, Vicolo dell'Osservatorio 3, I--35122, Padova, Italy}
\affiliation{INFN-Padova, Via Marzolo 8, I--35131 Padova, Italy}
\begin{abstract}
  We  investigate the impact of stellar rotation on the formation of black holes (BHs), by means of our population-synthesis code {\sc sevn}. Rotation affects the mass function of BHs in several ways. In massive metal-poor stars, fast rotation reduces the minimum zero-age main sequence (ZAMS) mass for a star to undergo pair instability and pulsational pair instability. Moreover, stellar winds are enhanced by rotation, peeling-off the entire hydrogen envelope. As a consequence of these two effects, the maximum BH mass we expect from the collapse a rotating metal-poor star is only $\sim{}45$ M$_\odot$, while the maximum mass of a BH born from a non-rotating star is $\sim{}60$ M$_\odot$. Furthermore, stellar rotation reduces the minimum ZAMS mass for a star to collapse into a BH from $\sim{}18-25$ M$_\odot$ to $\sim{}13-18$ M$_\odot$. Finally, we have investigated the impact of different core-collapse supernova (CCSN) prescriptions on our results. While the threshold value of compactness for direct collapse and the fallback efficiency strongly affect the minimum ZAMS mass for a star to collapse into a BH, the fraction of hydrogen envelope that can be accreted onto the final BH is the most important ingredient to determine the maximum BH mass. Our results confirm that the interplay between stellar rotation, CCSNe and pair instability plays a major role in shaping the BH mass spectrum.

\end{abstract}



\section{Introduction} \label{sec:intro}
The mass function of stellar black holes (BHs) is still an open question in astrophysics. Gravitational wave data are going to revolutionise our knowledge about BHs in the coming years: the first two observing runs of the LIGO--Virgo collaboration (LVC) led to the detection of ten binary BHs (\citealt{abbottO2,abbottO2popandrate}), few additional events were claimed by \cite{venumadhav2019} and \cite{zackay2019}, based on a different pipeline, and several new public triggers were announced during the third observing run of the LVC, which is still ongoing. This growing population of BHs complements the sample from dynamical mass measurements in nearby X-ray binaries \citep{oezel2010,farr2011} and will provide us with an unique opportunity to test BH formation models.



According to our current understanding, compact object masses are strictly related to the mass evolution and to the final fate of their progenitor stars. Massive stars ($\gtrsim{}30$ M$_\odot$) can lose a significant fraction of their initial mass by stellar winds, depending mostly (but not only) on their metallicity \citep{kudritzki1987,vink2001} and luminosity \citep{graefener2008,vink2011}. We expect that the final mass and the inner properties of a star at the onset of collapse have a strong impact on the final outcome of a core-collapse supernova (CCSN). If the final mass of the star is sufficiently large \citep{fryer1999,fryer2001} and the central compactness sufficiently high \citep{oconnor2011,ugliano2012}, a star might even avoid the final explosion and collapse to a BH quietly. Based on this reasoning, the maximum mass of BHs is predicted to depend on progenitor's metallicity, with metal-poor stars leaving more massive remnants than metal-rich ones \citep{heger2003,mapelli2009,mapelli2010,mapelli2013,belczynski2010,fryer2012,spera2015,spera2017}.

This basic framework is complicated by uncertainties on CCSN models (e.g. \citealt{janka2012,janka2017,foglizzo2015,sukhbold2016,pejcha2015,burrows2018,ebinger2019a,ebinger2019b}), by the existence of other explosion mechanisms, such as electron-capture supernovae \citep{nomoto1984,nomoto1987,jones2013}, pulsational pair instability supernovae (PPISNe) and pair instability supernovae (PISNe) \citep{fowler1964,barkat1967,woosley2007,woosley2017,woosley2019}, and by the complex physics of massive star evolution.

In particular, population-synthesis models used to investigate the mass function of (single and binary) BHs (e.g. \citealt{bethe1998, portegieszwart1998,  belczynski2002, belczynski2008, belczynski2010, mapelli2013,mennekens2014,spera2015,spera2017,eldridge2016,stevenson2017,mapelli2017,mapelli2018,giacobbo2018,giacobbo2018b,kruckow2018,spera2019,eldridge2019,mapelli2019,stevenson2019}) usually do not include stellar rotation among their ingredients. This might be a serious issue, because stellar rotation can dramatically affect  the evolution of the progenitor star \citep{limongi2018,dvorkin2018,groh2019}. Rotation has (at least) two competing effects on stellar evolution. It enhances chemical mixing \citep{meynet2005,ekstrom2012,chieffi2013,marchant2016, demink2016, mandel2016}, leading to the development of larger stellar cores, and at the same time enhances mass loss, quenching the final stellar mass (see e.g. \citealt{limongi2017} for a review).  Stars with He core $135\gtrsim{}M_{\rm He}/{\rm M}_\odot{}\gtrsim{}64$ are expected to undergo a PISN leaving no compact remnant.  Stars with $64\gtrsim{}M_{\rm He}/{\rm M}_\odot{}\gtrsim{}32$ experience enhanced mass loss because of pulsational pair instability.
  Since stellar rotation leads to the formation of more massive He cores, especially at low metallicity where winds are quenched, the minimum zero-age main sequence (ZAMS) mass for a rotating star to undergo PISN and PPISN can be significantly smaller than the minimum ZAMS mass for a non-rotating star.

Moreover, most population synthesis codes model the outcome of a CCSN explosion based on the carbon-oxygen mass of the progenitor star, following the prescriptions in \cite{fryer2012}, but hydrodynamical simulations of CCSNe suggest that this approach might be incomplete. For example, \cite{oconnor2011}  propose that the outcome of a CCSN, for a given equation of state, can be estimated, to first order, by the compactness of the stellar core at bounce, defined as
\begin{equation}\label{eq:compactness}
  \xi{}_{M}=\frac{M/{\rm M}_\odot}{R(M)/1000\,{}{\rm km}},
\end{equation}
where $R(M)$ is the radius that encloses a baryonic mass equal to $M$ at core bounce and $M$ is a given mass (usually $M=2.5$ M$_\odot$).

Here we present a new version of the population-synthesis code {\sc sevn} \citep{spera2015,spera2017,spera2019} in which we include stellar rotation by means of the {\sc franec} stellar evolution tracks \citep{limongi2000,chieffi2004,limongi2006,chieffi2013,limongi2018}. We discuss the impact of stellar rotation on compact-object mass. We also add a new simple prescription to include compactness and we compare the outcomes of CCSNe described by compactness with \cite{fryer2012} prescriptions.


\section{Methods} \label{sec:methods}

\subsection{SEVN}
           {\sc sevn}'s main difference with respect to most population synthesis codes is the approach to stellar evolution \citep{spera2015,spera2017,spera2019}. While the vast majority of population synthesis codes implements stellar evolution through the polynomial fitting formulas initially derived by \cite{hurley2000}, {\sc sevn} describes stellar evolution through look-up tables, obtained from stellar evolution tracks\footnote{{\sc combine} \citep{kruckow2018} is the only other binary population synthesis code (besides {\sc sevn}) that adopts look-up tables and has been used to study binary compact objects.}. The look-up tables contain information on star mass and core mass, star radius and core radius, stellar metallicity and evolutionary stages. Currently, the default tables are derived from the {\sc parsec} stellar evolution tracks \citep{bressan2012,tang2014,chen2015,marigo2017}. In this work, we describe the implementation of new tables derived from {\sc franec} (see the next section). The interpolation algorithm adopted in {\sc sevn} is already described in \cite{spera2017} and \cite{spera2019}. The main advantage of using look-up tables with respect to polynomial fitting formulas is that stellar evolution in {\sc sevn} can be updated very easily by changing the current set of look-up tables with a new one, while polynomial fitting formulas are bound to the stellar evolution model they were extracted from. 

           Binary evolution is implemented in {\sc sevn} following the prescriptions by \cite{hurley2002}. We include a treatment of tides, decay by gravitational-wave emission, mass transfer and common envelope as already discussed in \cite{spera2019}. The main novelty with respect to \cite{hurley2002} consists in the description of common envelope and stellar mergers. Thanks to the interpolation algorithm, the mass and the stellar type of the outcome of a common envelope or a stellar merger are  derived from the look-up tables directly, without the need for a collision matrix or other fitting formulas.

      Here below, we describe the new tables derived from {\sc franec} and the updates to the description of CCSN outcomes in {\sc sevn}.

\subsection{{\sc franec} stellar evolution tracks}
The stellar models adopted in this paper  have been computed by means of the latest release of the {\sc franec} code. Here, we summarize their main features, while  we refer to \cite{limongi2018} for a full description of the models and the code\footnote{The main properties of these models, together to their final yields, are available at the webpage \url{http://orfeo.iaps.inaf.it}. More specific details about the models may be provided upon request.}. 
The initial masses are 13, 15, 20, 25, 30, 40, 60, 80 and $\rm 120~M_\odot$, the initial metallicities are [Fe/H]$=0, -1, -2, -3$, and the initial equatorial rotation velocities are  0, 150 and 300 km s$^{-1}$.
We adopt the solar composition from \citet{agss09}, corresponding to a total heavy element mass fraction of $\rm Z_\odot\sim{}0.0135$. At metallicities lower than solar we consider a scaled solar distribution with the exception of  C, O, Mg, Si, S, Ar, Ca, and Ti, for which we assume 
 [C/Fe]=0.18, [O/Fe]=0.47, [Mg/Fe]=0.27, [Si/Fe]=0.37, [S/Fe]=0.35, [Ar/Fe]=0.35, [Ca/Fe]=0.33, [Ti/Fe]=0.23, consistent with the observations of unevolved metal-poor stars  \citep{cayreletal04,spiteetal05}. As a consequence, the total metallicities corresponding to [Fe/H]$=-1,-2,-3$ are $\rm Z\sim{}3\times 10^{-3},~3\times 10^{-4},~3\times 10^{-5}$, respectively.
The initial velocities were chosen to roughly span the range of observed values \citep{duftonetal06,hunteretal08,ram17}.

The nuclear network, fully coupled to the equations for the stellar structure as well as to the various kinds of mixing, includes
335 isotopes in total, from H to $^{209}$Bi, linked by more than 3000 nuclear reactions. This network is well suited  to properly follow all the stable and explosive nuclear burning stages of massive stars.

Mass loss is taken into account following different prescriptions for the various evolutionary stages, e.g.,
\cite{vink2000,vink2001} for the blue supergiant phase ($\rm T_{eff}>12000~K$), 
\citet{dejager88} for the red supergiant phase ($\rm T_{eff}<12000~K$) and \citet{nl00} for the Wolf-Rayet phase. 
The dust driven wind occurring during the red supergiant phase has been included 
following the prescriptions of \citet{vanloonetal05}. Mass loss is enhanced, in rotating models, 
according to \cite{hlw00}. When the star approaches the Eddington limit, mass loss is modeled as described in \cite{limongi2018}.

Rotation is treated as described in \citet{chieffi2013} and \cite{limongi2018}. Two main rotation driven instabilities are taken into account, i.e., 
meridional circulation and turbulent shear. The efficiency of the mixing induced by these two phenomena has been 
calibrated by requiring the fit to a subset of stars 
(taken from the LMC samples of the FLAMES survey, \citealt{untetal09}) for which both the surface N abundance and the 
projected rotation velocity are available.


\subsection{Core-collapse supernovae (CCSNe)}\label{sec:SNe}

{\sc sevn} includes five different models to describe the outcome of CCSNe: the rapid and delayed models presented in \cite{fryer2012}, the prescriptions adopted in {\sc startrack} \citep{belczynski2008}, the compactness criterion \citep{oconnor2011} and the two-parameter criterion by \cite{ertl2016}. The first three models depend only on the carbon-oxygen mass after carbon burning and on the pre-supernova mass of the star, the fourth model depends also on the compactness $\xi{}_{2.5}$, defined in equation~\ref{eq:compactness} (assuming $M=2.5$ M$_\odot$), while the fifth model depends on the enclosed mass at a dimensionless entropy per nucleon $s = 4$ ($M_4$) and the mass gradient at the same location ($\mu{}_4$). 

\subsubsection{Compactness model}\label{sec:compactness}
In the previous version of {\sc sevn}, the criterion based on compactness and the two-parameter criterion were implemented in a non-self-consistent way, because the table of compactness $\xi{}_{2.5}$ and that of $M_4$ and $\mu{}_4$ were calculated through the {\sc mesa} code \citep{paxton2011,paxton2013,paxton2015}, while stellar evolution was derived from {\sc parsec}. Here, we update the treatment of compactness in a self-consistent way. In fact, compactness can be calculated directly from {\sc franec} models, because they are evolved up to the onset of core collapse\footnote{\cite{oconnor2011} adopt compactness at bounce, but \cite{ugliano2012} show that compactness at the onset of collapse is consistent with compactness at bounce and is much easier to estimate. 
  Hereafter, we refer to compactness at the onset of collapse.}. 

\begin{figure}
\fig{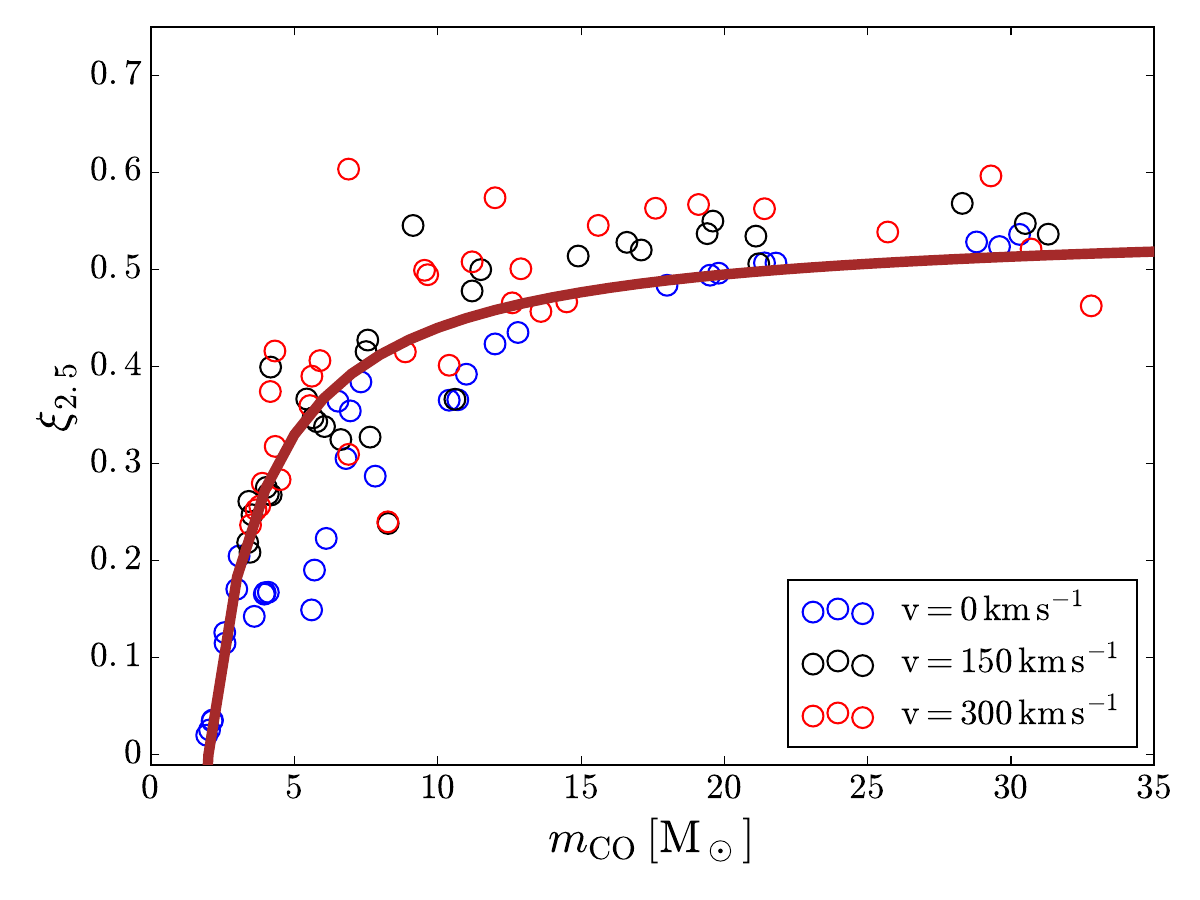}{0.4\textwidth}{}
\caption{Compactness $\xi{}_{2.5}$ as a function of the carbon-oxygen core mass ($m_{\rm CO}$) at the onset of collapse for the {\sc franec} evolutionary tracks with rotation $v=0,$ 150 and 300 km s$^{-1}$ (blue, black and red circles, respectively). The dark red line overlaid to the data is the fit described in equation~\ref{eq:fitcomp}.
\label{fig:fitcomp} }
\end{figure}

\cite{limongi2018} have shown that there is a strong correlation between compactness and carbon-oxygen mass at the onset of collapse (see their Figure~21) and this correlation is not significantly affected by stellar rotation. 
Thus, in our new version of {\sc sevn}, we interpolate compactness among stellar models by using the following fitting formula:
\begin{equation}\label{eq:fitcomp}
  \xi{}_{2.5}=a+b\,{}\left(\frac{m_{\rm CO}}{1\,{}{\rm M}_\odot}\right)^c,
\end{equation}
where $a=0.55$, $b=-1.1$, $c=-1.0$. 
Figure~\ref{fig:fitcomp} shows the fit reported in equation~\ref{eq:fitcomp} overlaid to the data of {\sc franec}.

\cite{oconnor2011} suggest that progenitors with $\xi{}_{2.5}>0.45$ most likely form BHs without explosion, while \cite{horiuchi2014} suggest a lower threshold value ($\xi{}_{2.5}\gtrsim{}0.2$). In this work, we adopt $\xi_{2.5}=0.3$ as threshold (unless explicitly stated otherwise) and we simply assume that progenitors with $\xi{}_{2.5}\leq{}0.3$  form a neutron star (NS) by CCSN explosion, while progenitors with $\xi{}_{2.5}>0.3$ form a BH by direct collapse.

Several recent papers claim that $\xi{}_{2.5}$ does not show a monotonic trend with the CO core (\citealt{sukhbold2018} and references therein), but rather has a complicated trend, with several localized branches and multivalued solutions. This result 
is still a matter of debate. We are studying  this problem in detail and will discuss our results in a forthcoming paper. For this reason, and for the purposes of the present paper, here we adopt a conservative approach based on the results presented in \cite{limongi2018}. 

 The compactness criterion allows us to discriminate between the formation of a NS (if the progenitor explodes) and that of a BH (if the progenitor collapses directly). When the progenitor explodes leaving a NS, the mass of the NS is assigned randomly, following a Gaussian distribution with mean $\langle{}m_{\rm NS}\rangle{}=1.33$ M$_\odot$ and dispersion $\sigma{}_{\rm NS}=0.09$ M$_\odot$, based on the distribution of observed NSs in binary NS systems \citep{oezel2016}. 

 When the progenitor undergoes a direct collapse,  the mass of the BH is derived as
\begin{equation}\label{eq:BHmass}
    m_{\rm BH}=m_{\rm He}+f_{\rm H}\,{}(m_{\rm fin}-m_{\rm He}),
\end{equation}
where $m_{\rm fin}$ and $m_{\rm He}$ are the total mass and the He core mass of the star at the onset of collapse, respectively (the He core, by definition, includes also heavier elements inside the He core radius), while $f_{\rm H}$ is a free parameter which can assume values from 0 to 1. The presence of $f_{\rm H}$ accounts for the uncertainty about the collapse of the H envelope (if the progenitor star retains a H envelope to the very end). Some studies (e.g. \citealt{nadezhin1980,lovegrove2013,sukhbold2016,fernandez2018}) stress that is quite unlikely that the H envelope collapses entirely, even during a direct collapse, because it is loosely bound. In the following, we consider the two extreme cases in which $f_{\rm H}=0$ (the H envelope is completely lost) and $f_{\rm H}=0.9$ (90~\% of the H envelope collapses). Equation~\ref{eq:BHmass} is a toy model and does not intend to capture the complex physics of direct collapse. However, if we consider the two extreme cases with $f_{\rm H}=0$ and $f_{\rm H}=0.9$, we are able to bracket the main uncertainties on direct collapse.

In the compactness model, we assume that the efficiency of fallback is negligible, following recent hydrodynamical simulations (e.g. \citealt{ertl2016}).

\subsubsection{Rapid model}
               In this work, we compare  the new compactness criterion implemented in {\sc{sevn}} with the rapid CCSN model by \cite{fryer2012}, which assumes that the explosion occurs $<250$ ms after bounce. In the rapid model, the mass of the compact object is $m_{\rm rem}=m_{\rm proto}+m_{\rm fb}$, where $m_{\rm proto}=1$~M$_\odot$ is the mass of the proto-compact object and $m_{\rm fb}=f_{\rm fb}\,{}(m_{\rm fin}-m_{\rm proto})$ is the mass accreted by fallback. In the previous expression, $f_{\rm fb}$ is the fractional fallback parameter, defined as in \cite{fryer2012}.

                In the rapid CCSN formalism, the maximum NS mass is 2 M$_\odot$, while the minimum BH mass is 5 M$_\odot$. This result strongly depends on the assumptions about fallback. In contrast, our compactness-based model cannot predict a maximum NS mass, because the mass of the NS is derived from an observational distribution \citep{oezel2016}.

                We stress that none of the prescriptions currently adopted in the literature to infer the mass of compact objects (including the rapid model and the compactness-based models adopted in this work) is sufficient to capture the complexity of CCSN physics (see e.g. \citealt{burrows2018,burrows2019,vartanyan2019}). The aim of our study is to compare different CCSN prescriptions and to quantify the uncertainties on BH mass spectrum that arise from a different choice of these simplified prescriptions.

\subsection{PPISNe and PISNe}
{\sc sevn} includes a treatment for PISNe and PPISNe as described in \cite{spera2017}, based on the results of \cite{woosley2017}. In particular, if the He core mass is $135\ge{}m_{\rm He}/{\rm M}_\odot\ge{}64$, the star undergoes a PISN and leaves no compact object. If the He core mass is $64>m_{\rm He}/{\rm M}_\odot\ge{}32$, the star undergoes pulsational pair instability and the final mass of the compact object is calculated as $m_{\rm rem}=\alpha_{\rm P}\,{}m_{\rm no\,{}PPI}$, where $m_{\rm no\,{}PPI}$ is the mass of the compact object we would have obtained if we had not included pulsational pair instability in our analysis (just CCSN) and $\alpha_{\rm P}$ is a fitting parameter described in Appendix~\ref{sec:appendix}. 

\section{Results} \label{sec:results}

\begin{figure*}
\gridline{\fig{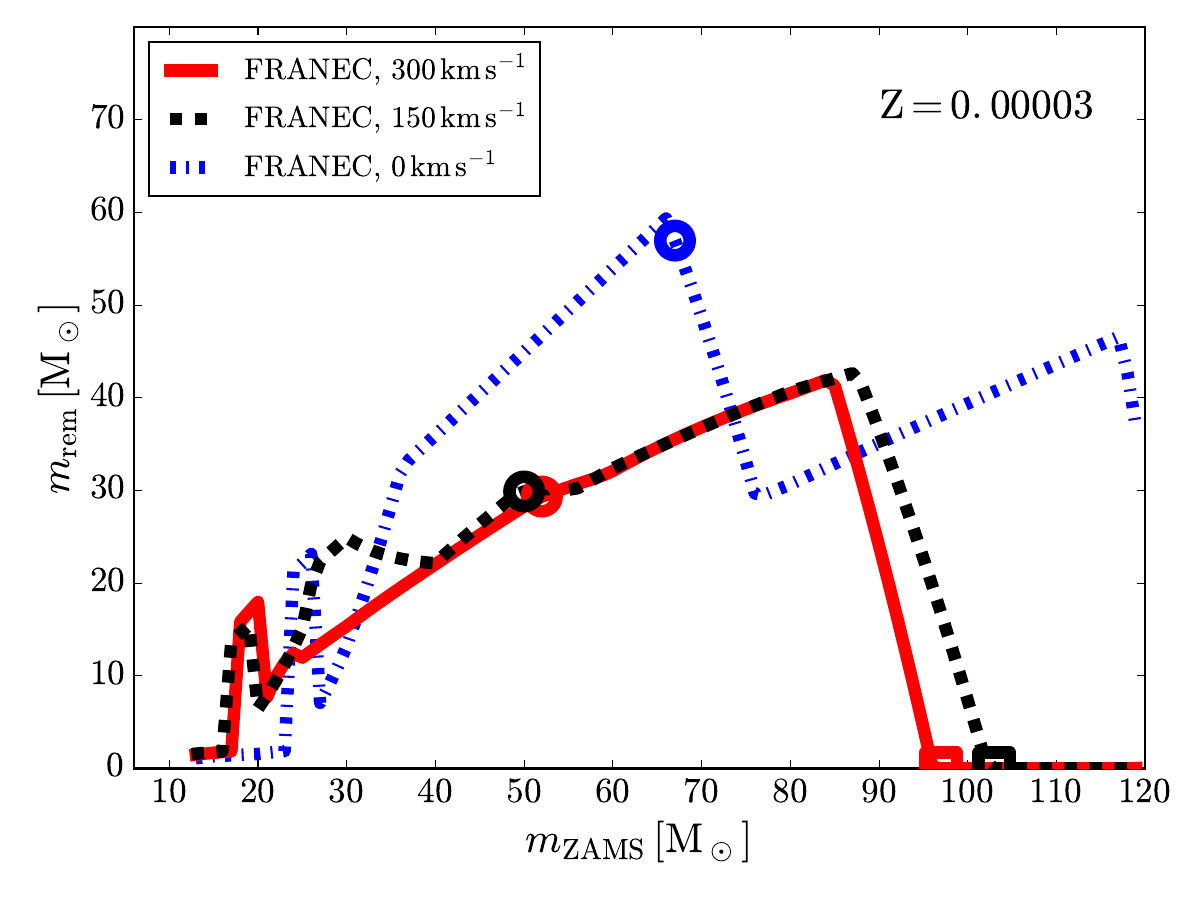}{0.5\textwidth}{}
          \fig{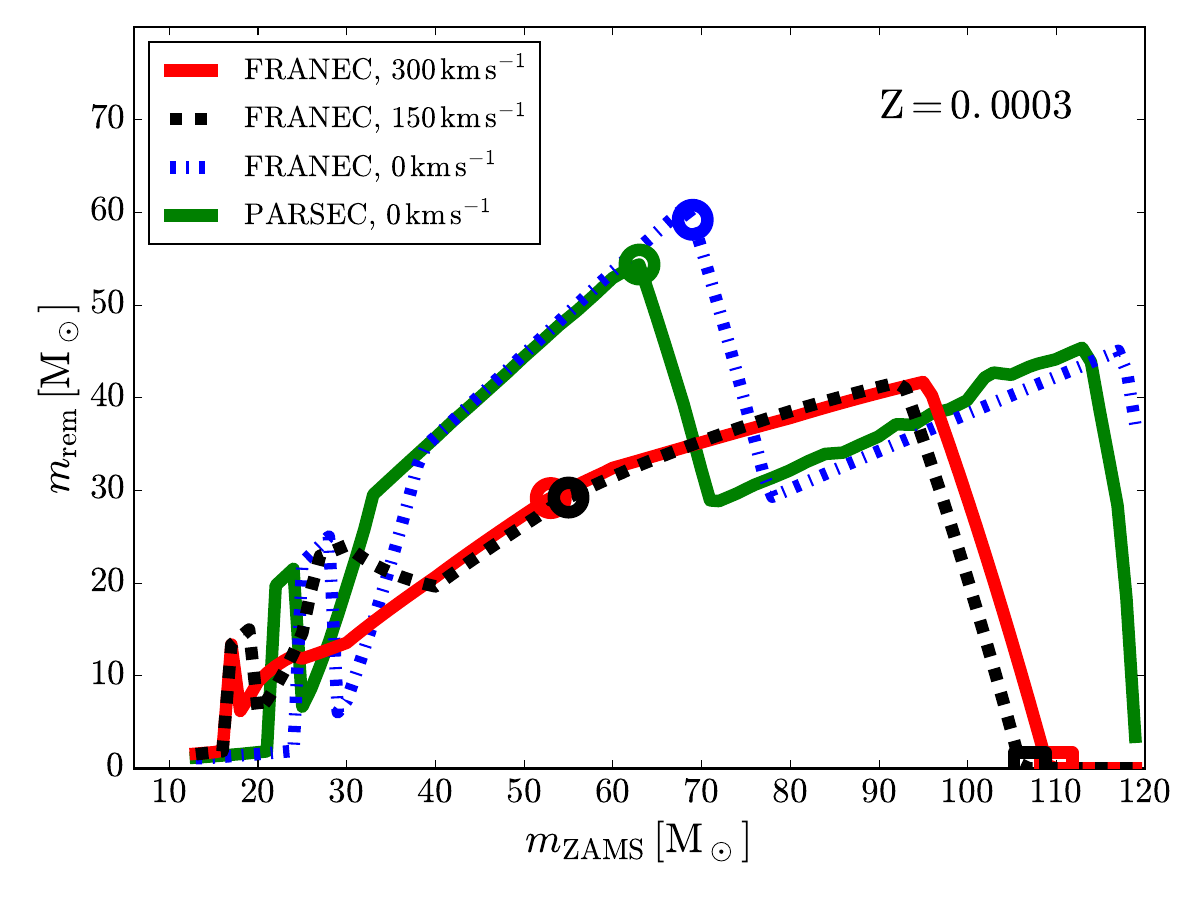}{0.5\textwidth}{}
          }
\gridline{\fig{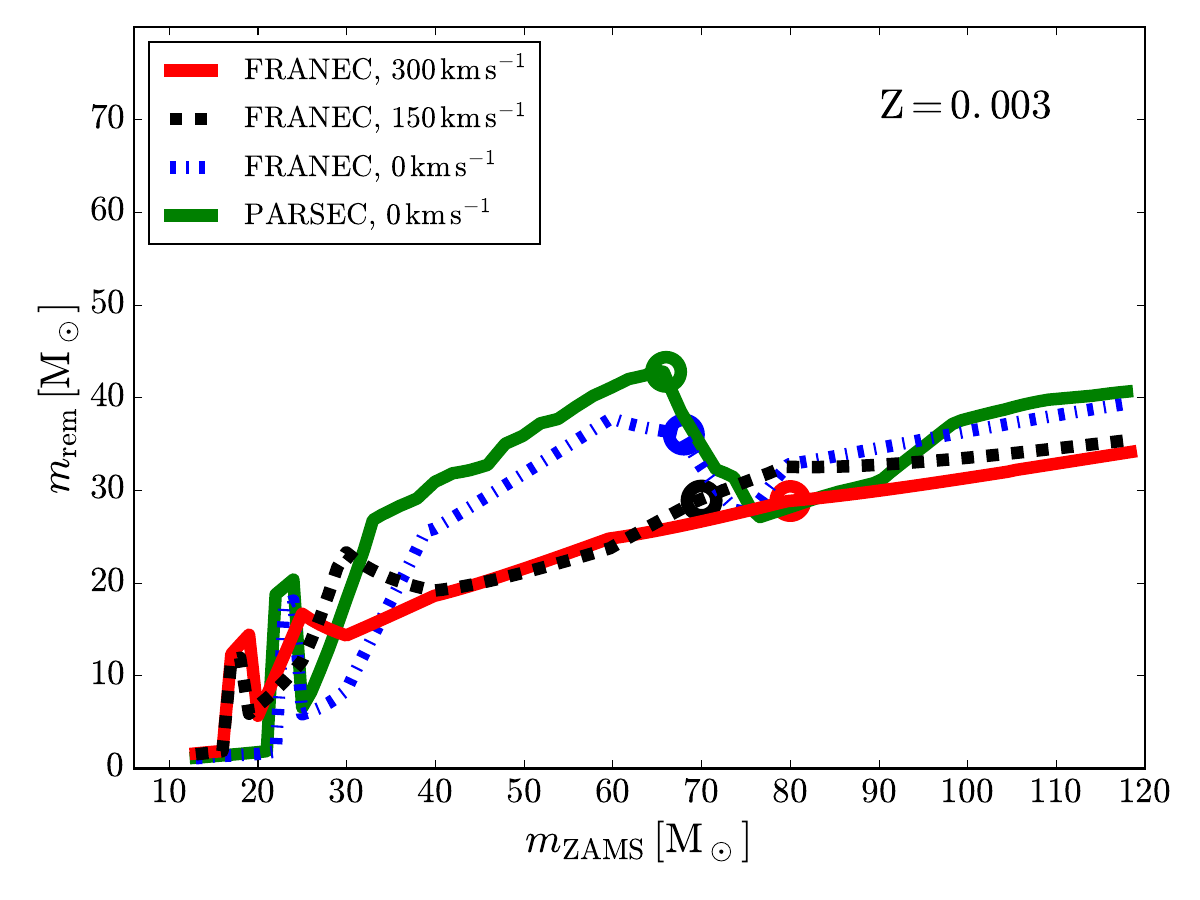}{0.5\textwidth}{}
          \fig{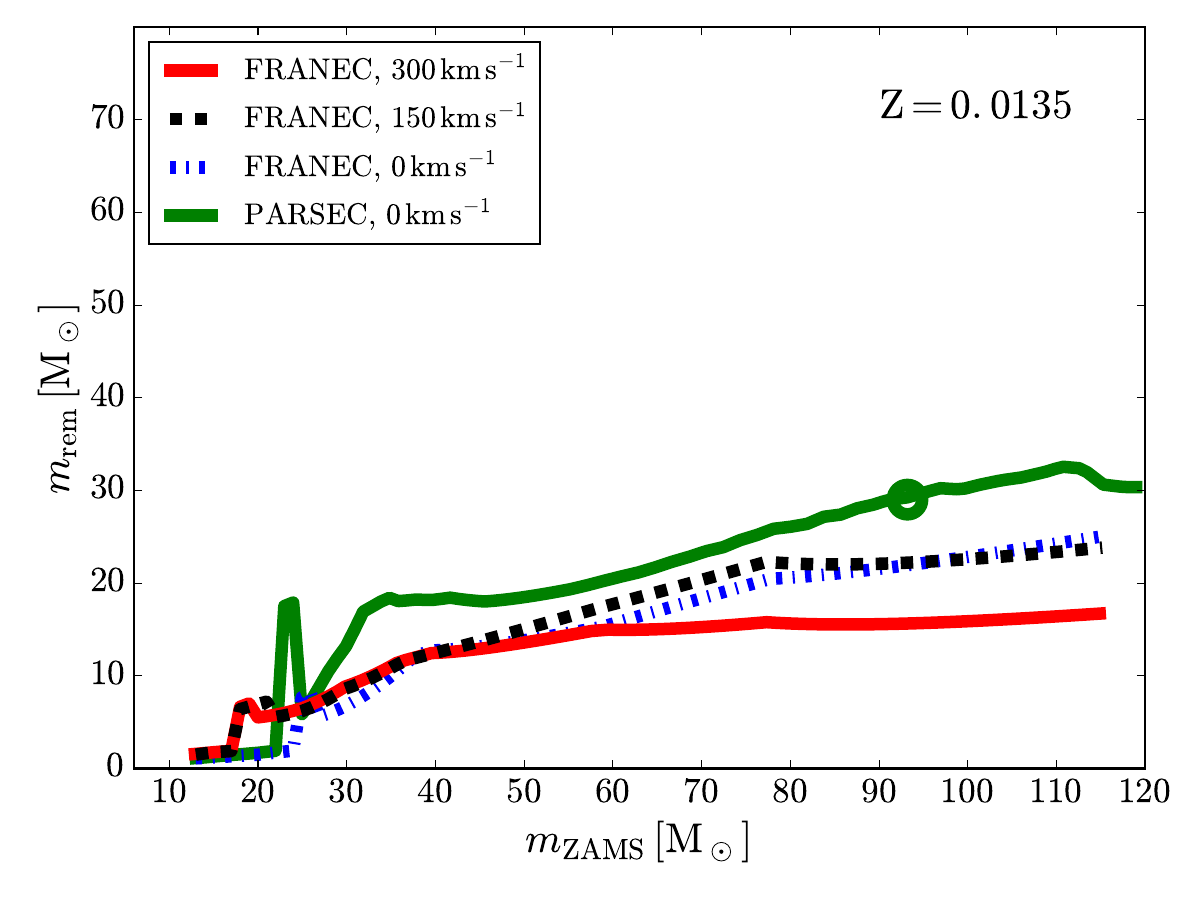}{0.5\textwidth}{} 
          }
\caption{Estimated mass of the compact object ($m_{\rm rem}$) as a function of the zero-age main sequence (ZAMS) mass of the progenitor star ($m_{\rm ZAMS}$). The outcome of CCSNe is described by the rapid model \citep{fryer2012}. From top to bottom and from left to right: $Z=0.00003$, 0.0003, 0.003 and 0.0135. Red  solid line: stellar evolution is described by {\sc franec} \citep{limongi2018} with initial equatorial rotation speed $v=300$ km s$^{-1}$. Black  dashed line: {\sc franec} \citep{limongi2018} with  $v=150$~km~s$^{-1}$. Blue  dot-dashed line: {\sc franec} \citep{limongi2018} with  $v=0$ km s$^{-1}$. Green  solid line: stellar evolution is described by {\sc parsec} \citep{bressan2012}. We do not have {\sc parsec} models with metallicity $Z=0.00003$. Open circles (squares): ZAMS mass at which the star develops a He core $m_{\rm He}=32$ M$_\odot$ ($m_{\rm He}=64$ M$_\odot$), corresponding to the minimum mass to undergo PPISN (PISN). 
\label{fig:rapid} }
\end{figure*}

\subsection{Impact of rotation on BH masses}\label{sec:rotation}
\begin{figure*}
\gridline{\fig{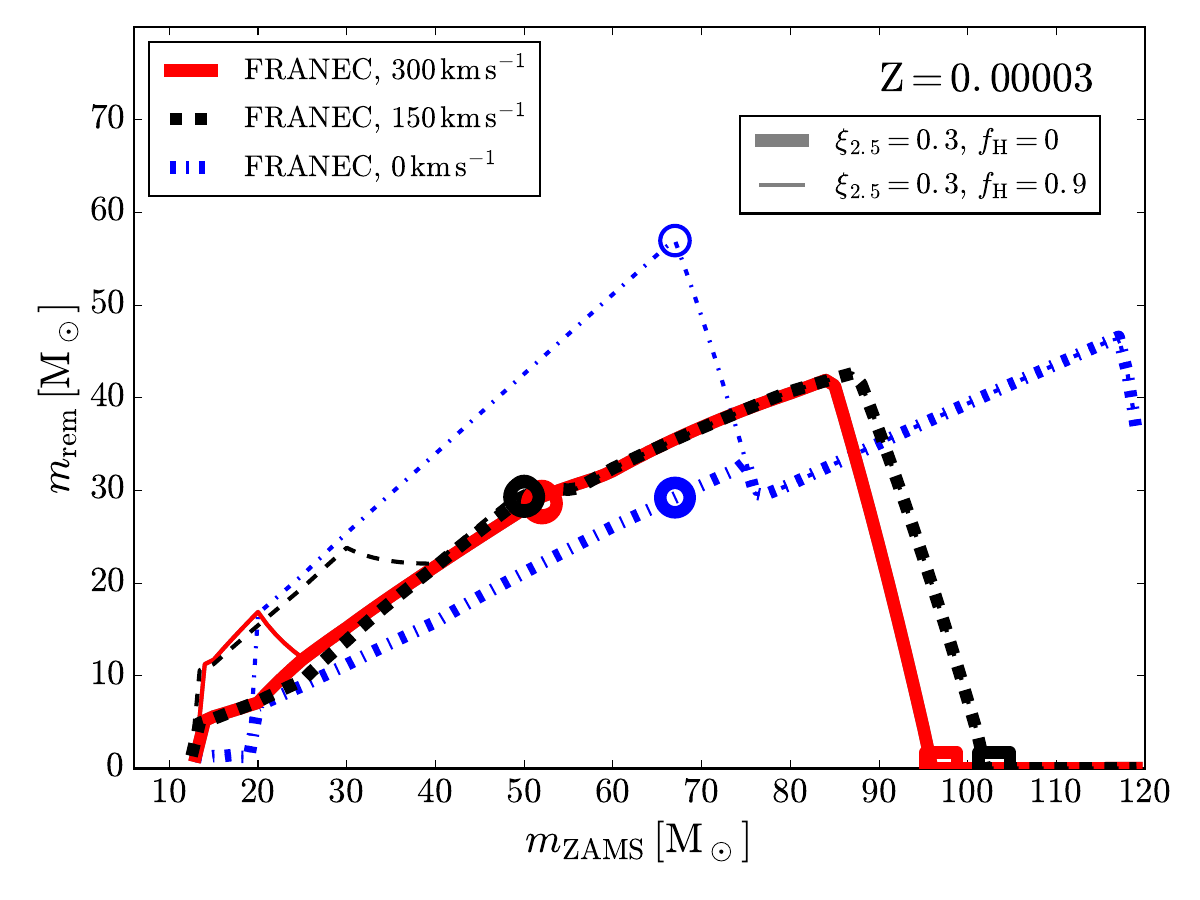}{0.5\textwidth}{}
          \fig{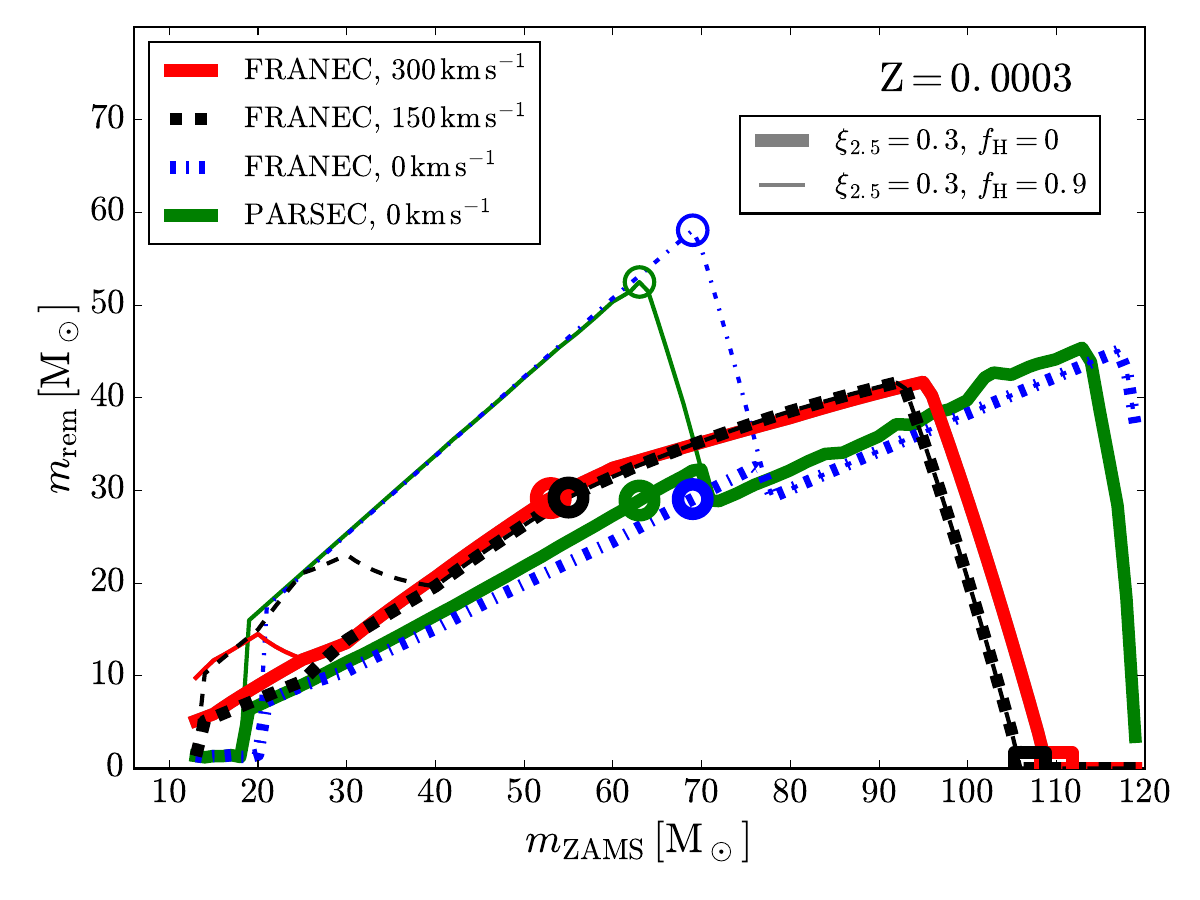}{0.5\textwidth}{}
          }
\gridline{\fig{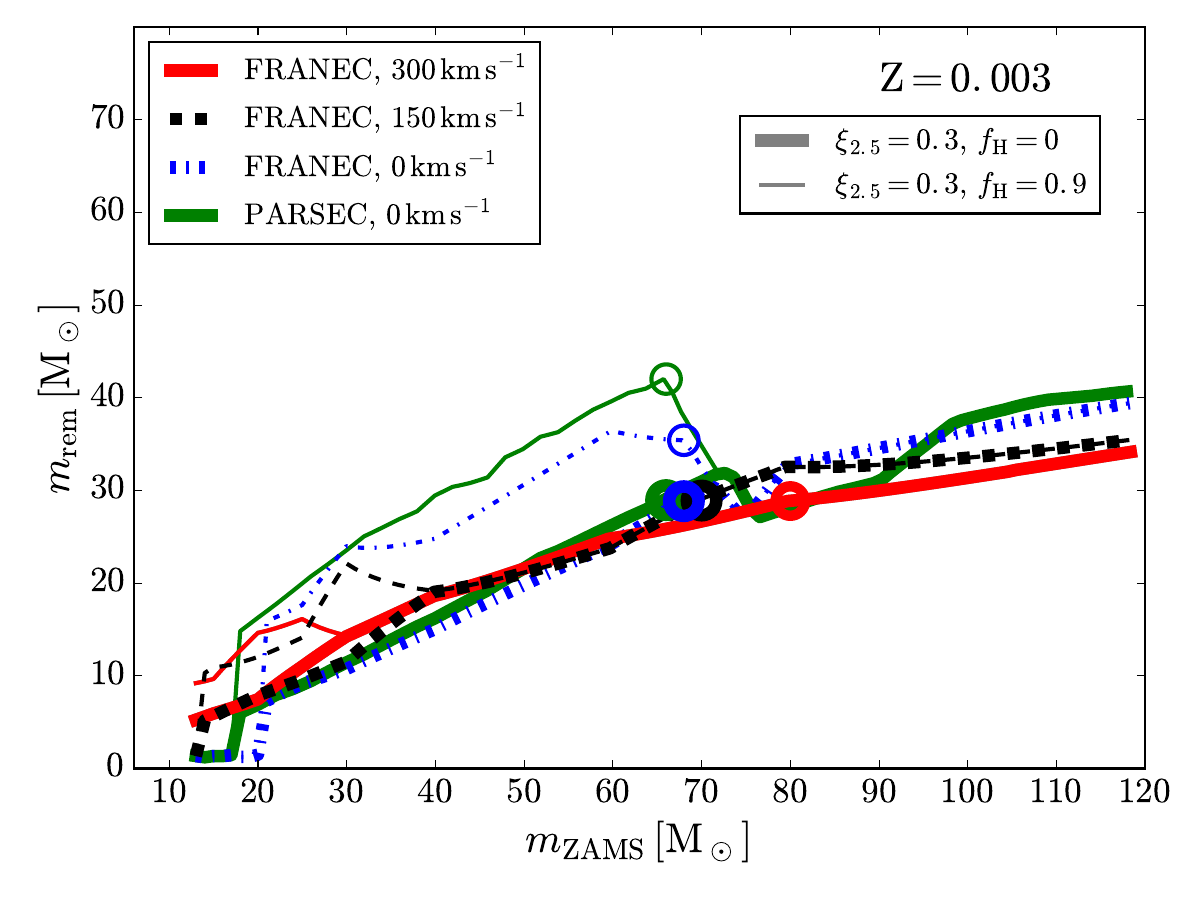}{0.5\textwidth}{}
          \fig{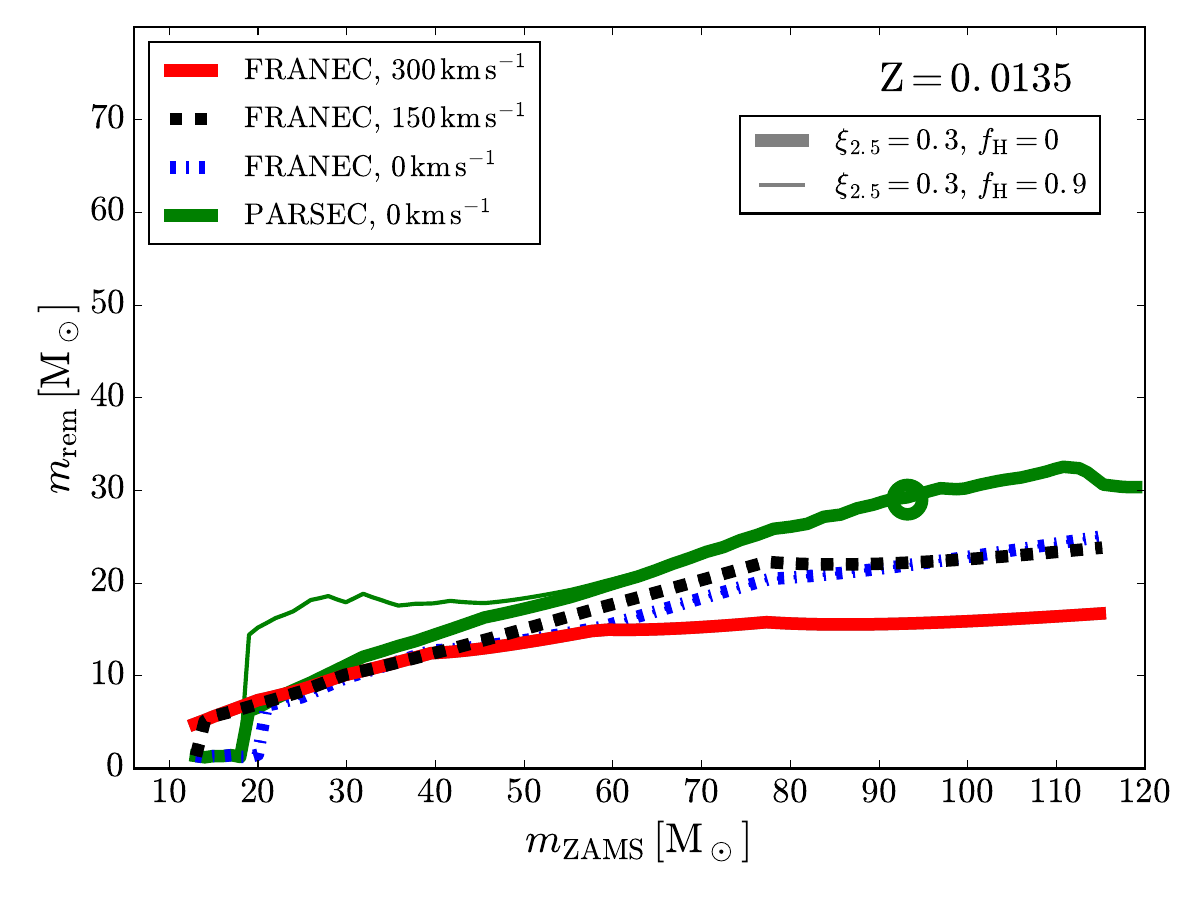}{0.5\textwidth}{} 
          }
\caption{Same as Figure~\ref{fig:rapid}, but CCSNe are described with the compactness criterion. Thick lines: we assume $\xi{}_{2.5}=0.3$ and $f_{\rm H}=0.0$; thin lines: we assume $\xi{}_{2.5}=0.3$ and $f_{\rm H}=0.9$ (see Section~\ref{sec:compactness}). If the thin lines are not visible, it means that they overlap with the thick lines perfectly. 
\label{fig:compac}}
\end{figure*}

Figures~\ref{fig:rapid} and \ref{fig:compac} show the mass of compact objects as a function of the ZAMS mass of their progenitor stars for different CCSN models (rapid, compactness with $\xi{}_{2.5}=0.3$ and $f_{\rm H}=0$, and compactness with $\xi{}_{2.5}=0.3$ and $f_{\rm H}=0.9$). We show the results we obtain with {\sc franec} stellar evolution tables for three initial  equatorial  velocities of the progenitor stars: $v=0$, 150 and 300 km s$^{-1}$. For comparison, we show also the results of {\sc parsec} stellar evolution tables with $v=0$ km s$^{-1}$. 

From these Figures it is apparent the strong impact of rotation on the minimum ZAMS mass for BH formation, regardless of progenitor's metallicity. The minimum progenitor mass to collapse to a BH is $m_{\rm ZAMS}\sim{}13-18$ M$_\odot$  for rotating stars and $m_{\rm ZAMS}\sim{}18-25$ M$_\odot$ for non-rotating stars (with a mild dependence on the CCSN model, see Table~\ref{tab:table1}). This happens because stars with $10\lesssim{}m_{\rm ZAMS}/{\rm M}_\odot{}\lesssim{}30$ are not particularly affected by stellar winds, regardless of their metallicity. Thus, angular momentum is not efficiently removed by mass loss and rotation has enough time to induce chemical mixing, leading to the growth of the stellar core. This shifts the threshold between explosion and direct collapse towards lower ZAMS masses.

Furthermore, Figures~\ref{fig:rapid} and \ref{fig:compac}  show that stellar rotation has a strong impact on the (pulsational) pair-instability window for metal-poor stars ($Z=0.0003,\,{}0.00003$), independent of the assumed CCSN model. The most metal-poor rotating models ($Z=0.0003,\,{}0.00003$) undergo PISN and PPISN at significantly lower ZAMS masses than the non-rotating models (e.g. $m_{\rm PPISN}\sim{} 50$ M$_\odot$ and $\sim{}70$ M$_\odot$ for rotating and non-rotating models, respectively, see Table~\ref{tab:table1}). Again, this happens because chemical mixing leads to significantly larger He cores in rotating metal-poor stars. We note that there are no significant differences between $v=150$ km s$^{-1}$ and $v=300$ km s$^{-1}$.

We now go through different metallicities, to discuss how the effect of stellar rotation changes with $Z$. 
In metal-poor stars ($Z\leq{}0.0003$), stellar winds are relatively inefficient over the entire mass spectrum, even for rotating stars. Thus, the main effect of rotation is always the enhancement of chemical mixing, leading to the growth of the stellar core. This has the two main consequences we discussed above, i.e. a smaller minimum ZAMS mass for BH formation and a smaller minimum ZAMS mass for PPISNe and PISNe.

In contrast, at intermediate metallicity ($Z=0.003$, approximately $1/5$ of the solar metallicity), the impact of rotation is different for stars with $m_{\rm ZAMS}\lesssim{}30$ M$_\odot$ and $m_{\rm ZAMS}\gtrsim{}30$ M$_\odot$. If $m_{\rm ZAMS}\lesssim{}30$ M$_\odot$, stellar winds are not particularly efficient, even in rotating models. Thus, rotating stars develop larger cores and end their life with higher compactness than non-rotating stars. The main consequence of this is that the minimum progenitor mass to collapse to a BH is smaller for rotating stars than for non-rotating stars. 
In contrast, if $m_{\rm ZAMS}\gtrsim{}30$ M$_\odot$, stellar winds are efficient at $Z=0.003$ and they are significantly enhanced by rotation. Because of enhanced mass loss, the He core of rotating stars tends to be smaller than the He core of non-rotating stars. As a consequence of this, at $Z=0.003$ the minimum ZAMS mass to enter the PPISN regime is slightly lower for non-rotating models ($m_{\rm PPISN}\sim{}66-68$~M$_\odot$  for $v=0$~km~s$^{-1}$, Table~\ref{tab:table1}) than for rotating models ($m_{\rm PPISN}\sim{}80$~M$_\odot$  for $v=300$ km s$^{-1}$, Table~\ref{tab:table1}), with an opposite behavior with respect to more metal-poor stars. Stars with $m_{\rm ZAMS}\leq{}120$ M$_\odot$ and $Z=0.003$ do not develop He cores $>64$ M$_\odot$, thus they do not enter the PISN regime.

Finally, metal-rich stars ($Z=0.0135\sim{}{\rm Z}_\odot$) with  $m_{\rm ZAMS}\leq{}30$ M$_\odot$ behave similarly to metal-poor stars: they are only mildly affected by mass loss; hence, rotating stars grow larger He cores than non-rotating stars, causing the minimum ZAMS mass for BH formation to shift to lower values in rotating models. In contrast, stellar winds are so efficient in metal-rich stars with $m_{\rm ZAMS}\gtrsim{}30$ M$_\odot$ that they do not enter either the PPISN or PISN window, regardless of their rotation speed (with the exception of the {\sc parsec} model, which undergoes PPISNe at $m_{\rm ZAMS}\gtrsim{}94$ M$_\odot$). At high $Z$, stellar rotation does not affect significantly the maximum BH mass, which is $\sim{}16-24$ M$_\odot$, regardless of the assumed CCSN model.


\begin{figure*}
\gridline{\fig{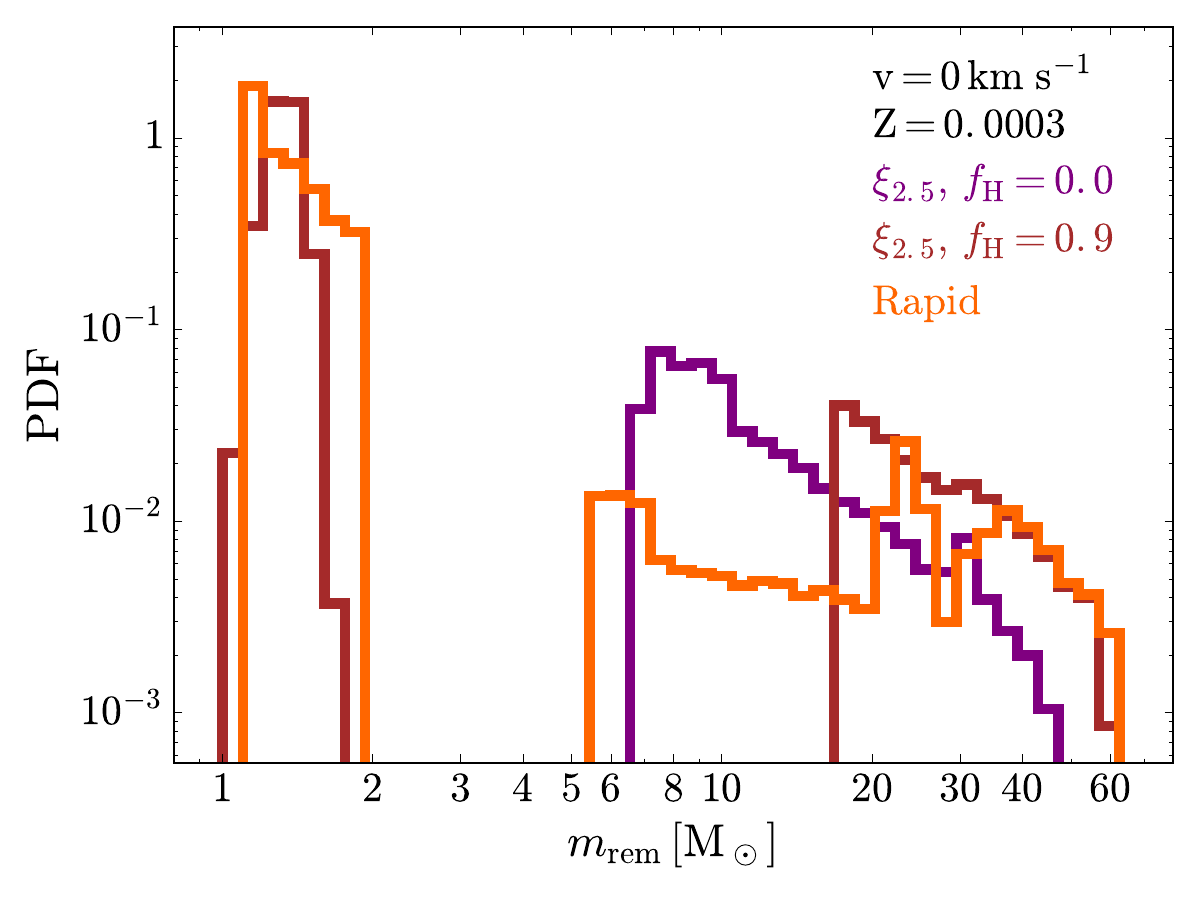}{0.33\textwidth}{}
  \fig{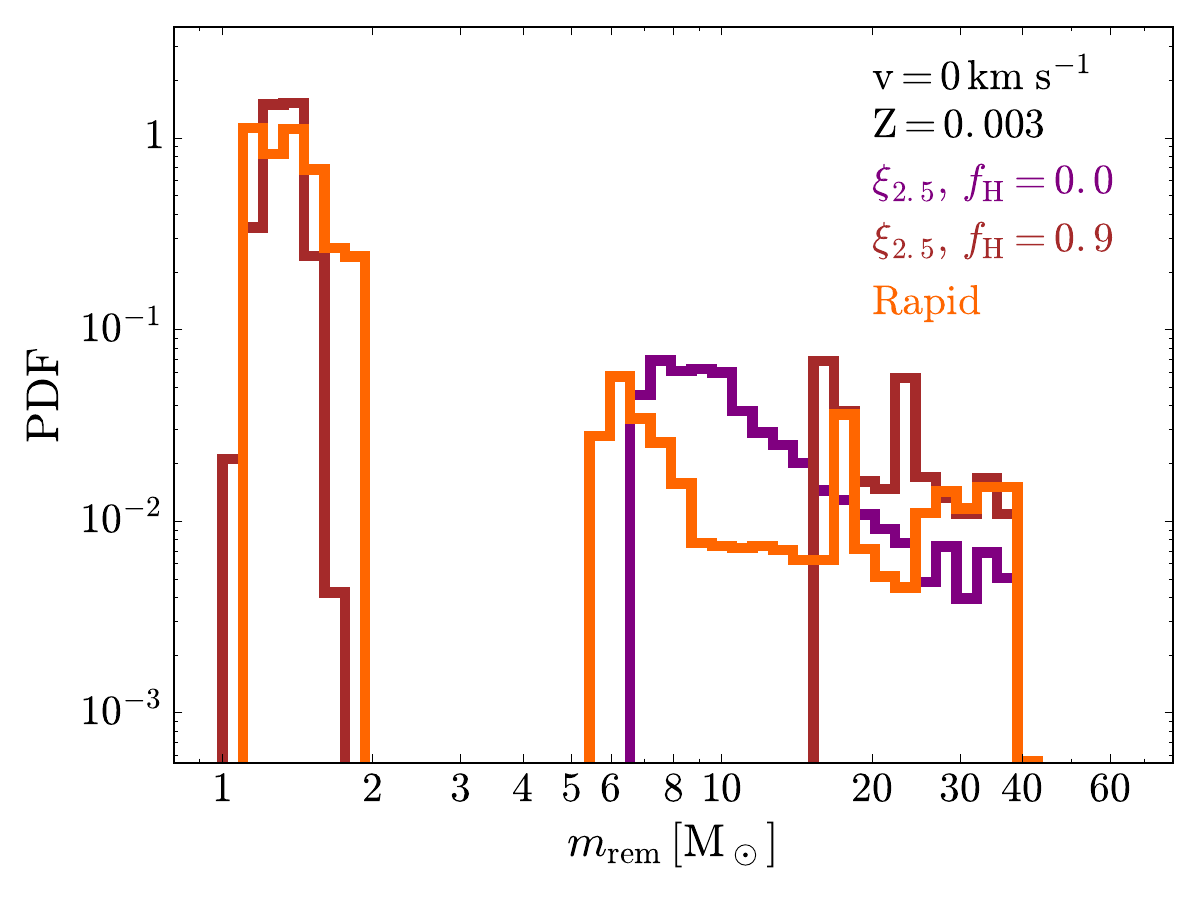}{0.33\textwidth}{}
            \fig{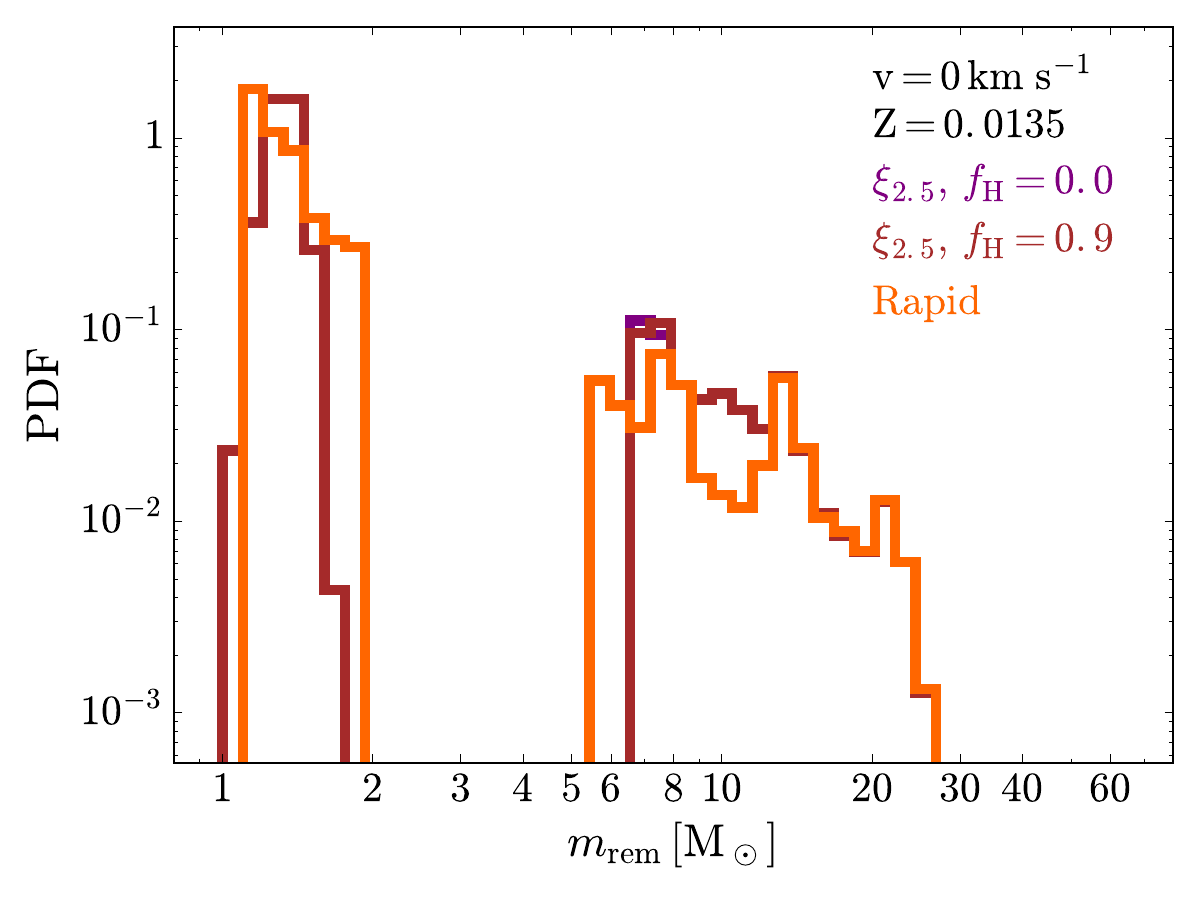}{0.33\textwidth}{}
}
\gridline{\fig{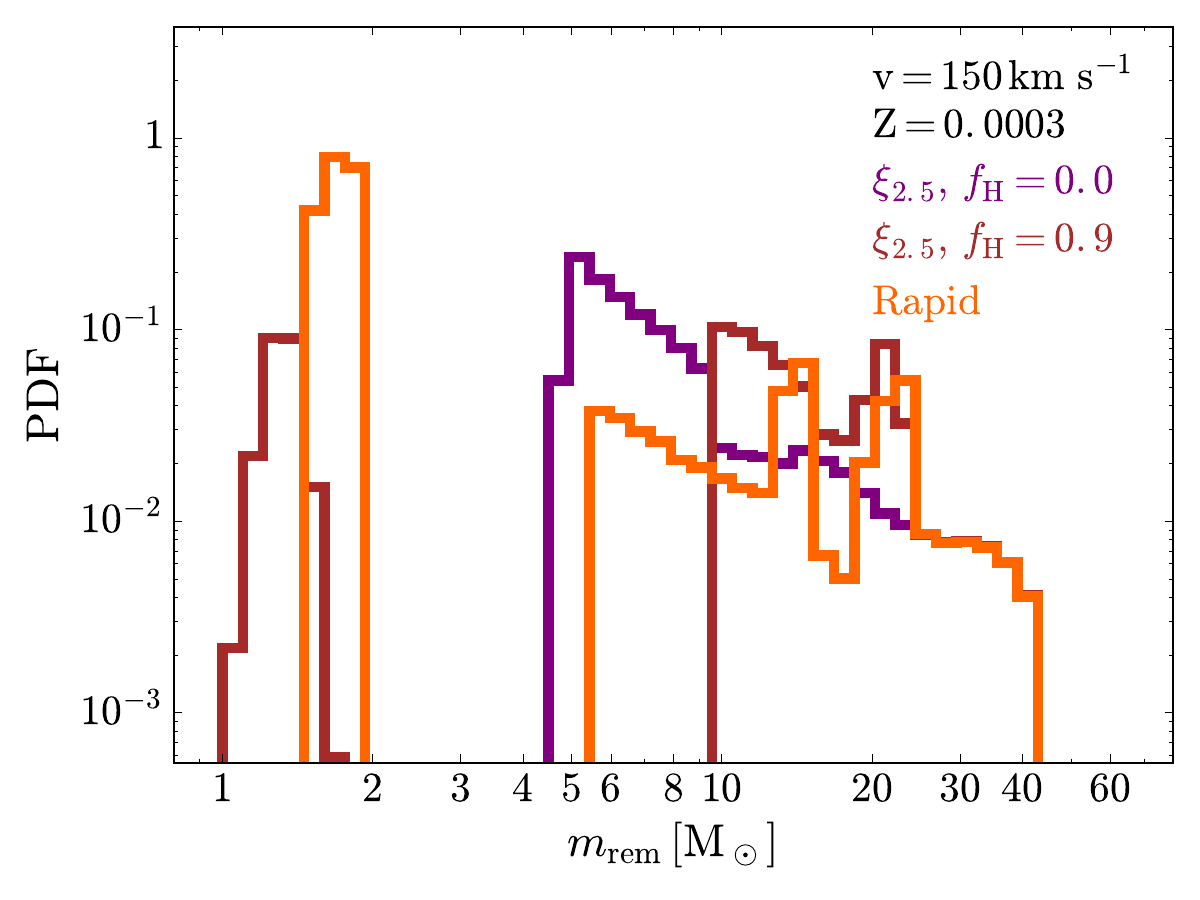}{0.33\textwidth}{}
  \fig{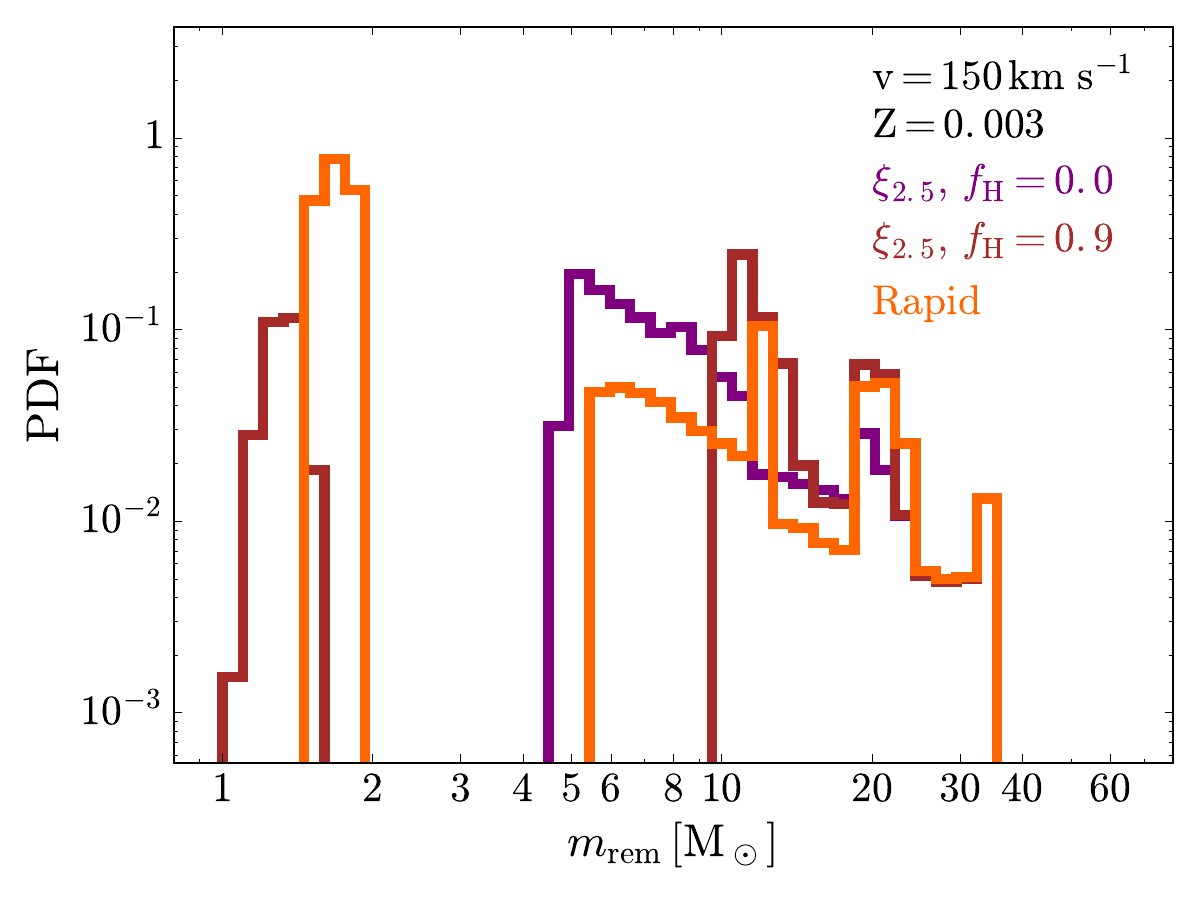}{0.33\textwidth}{}
            \fig{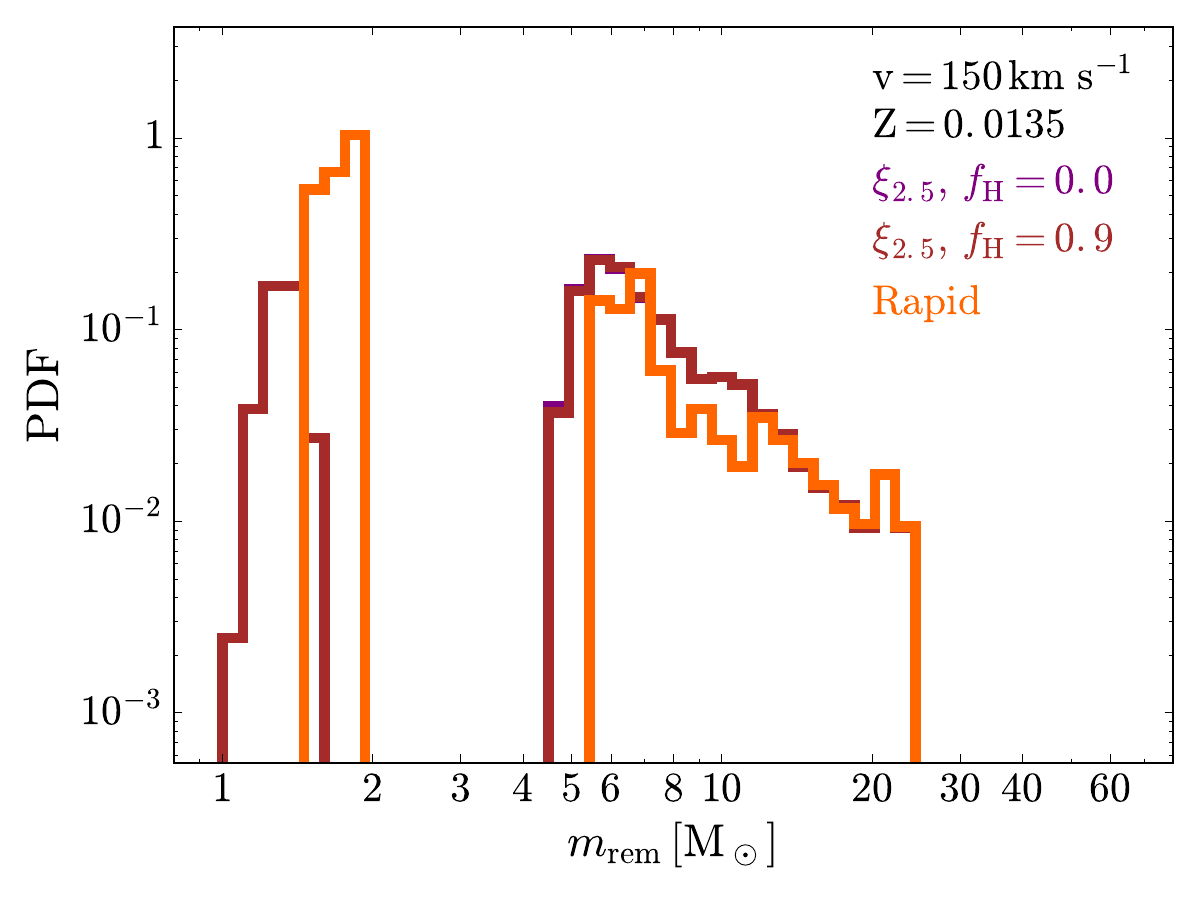}{0.33\textwidth}{}
}
\gridline{\fig{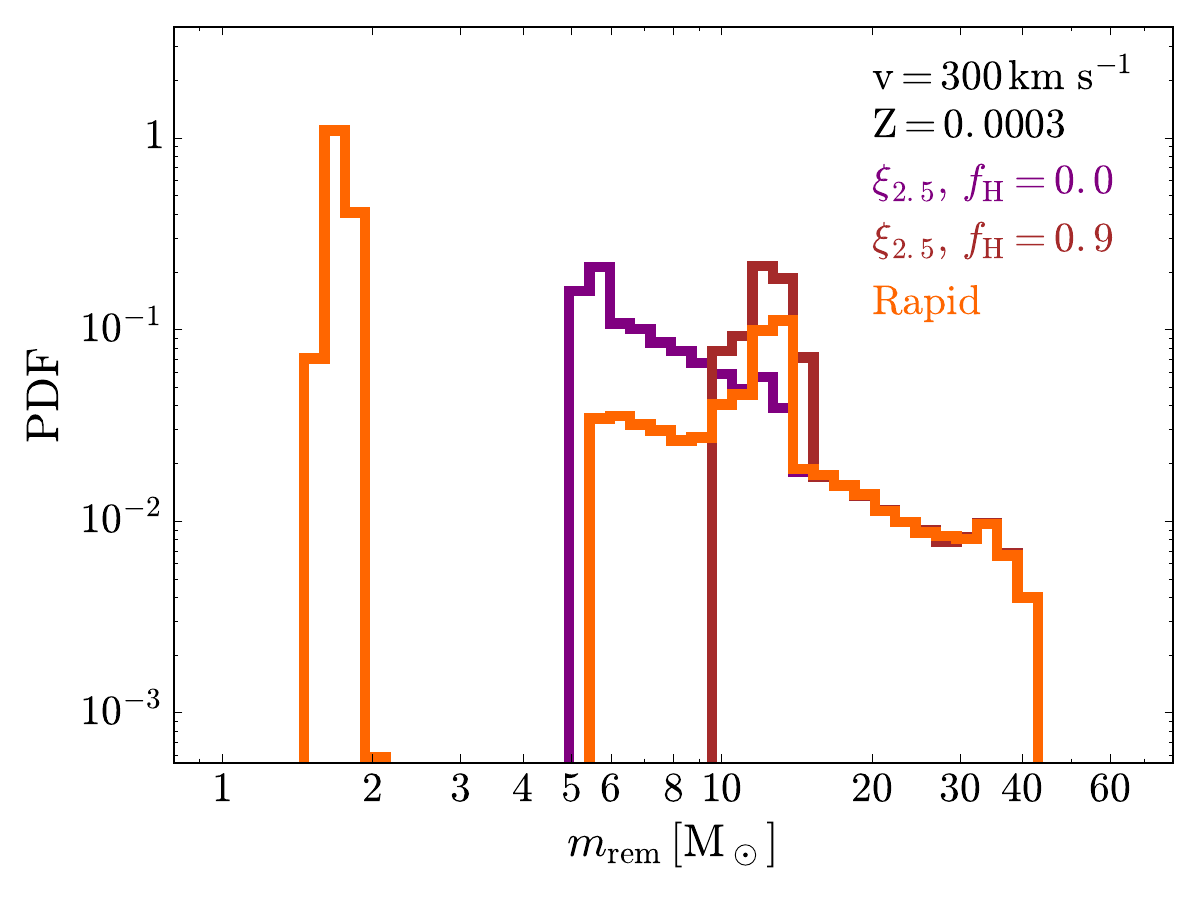}{0.33\textwidth}{}
  \fig{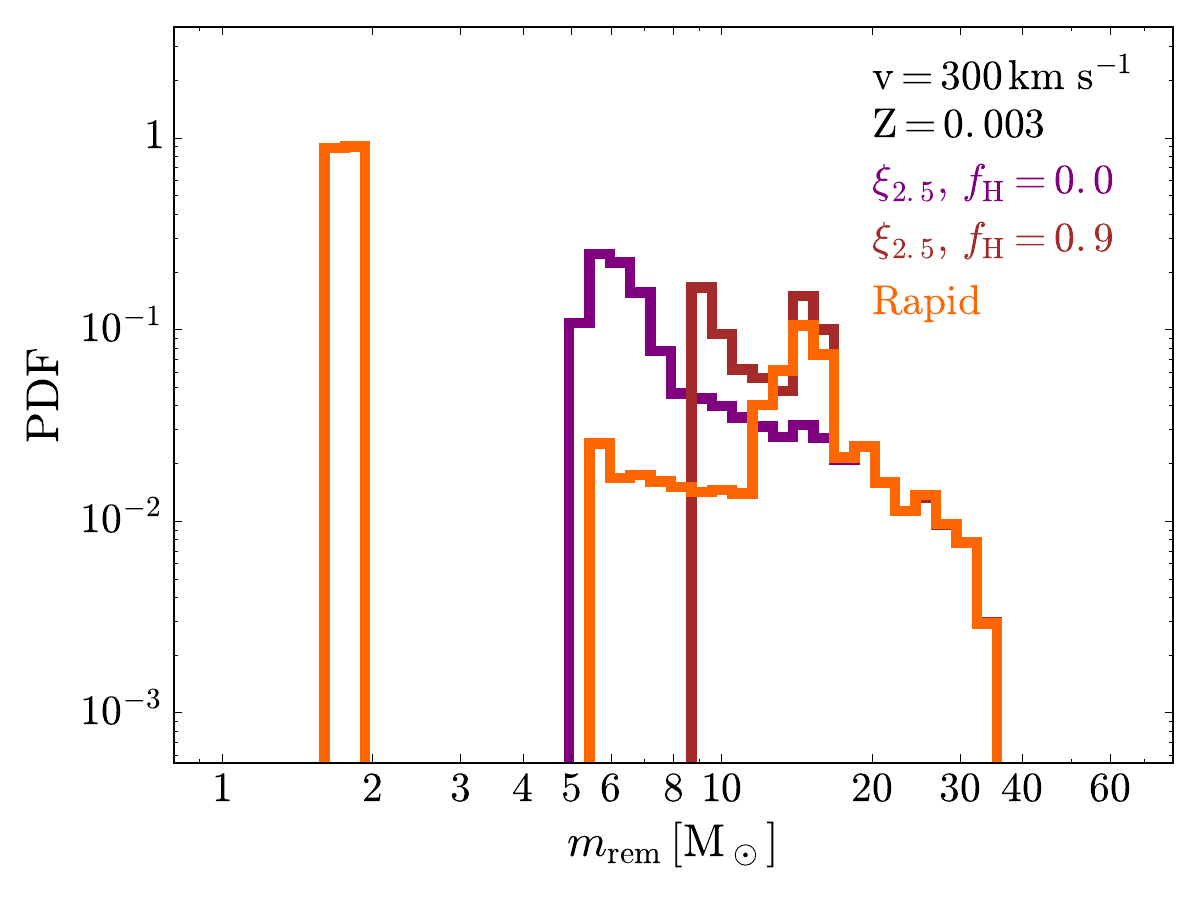}{0.33\textwidth}{}
            \fig{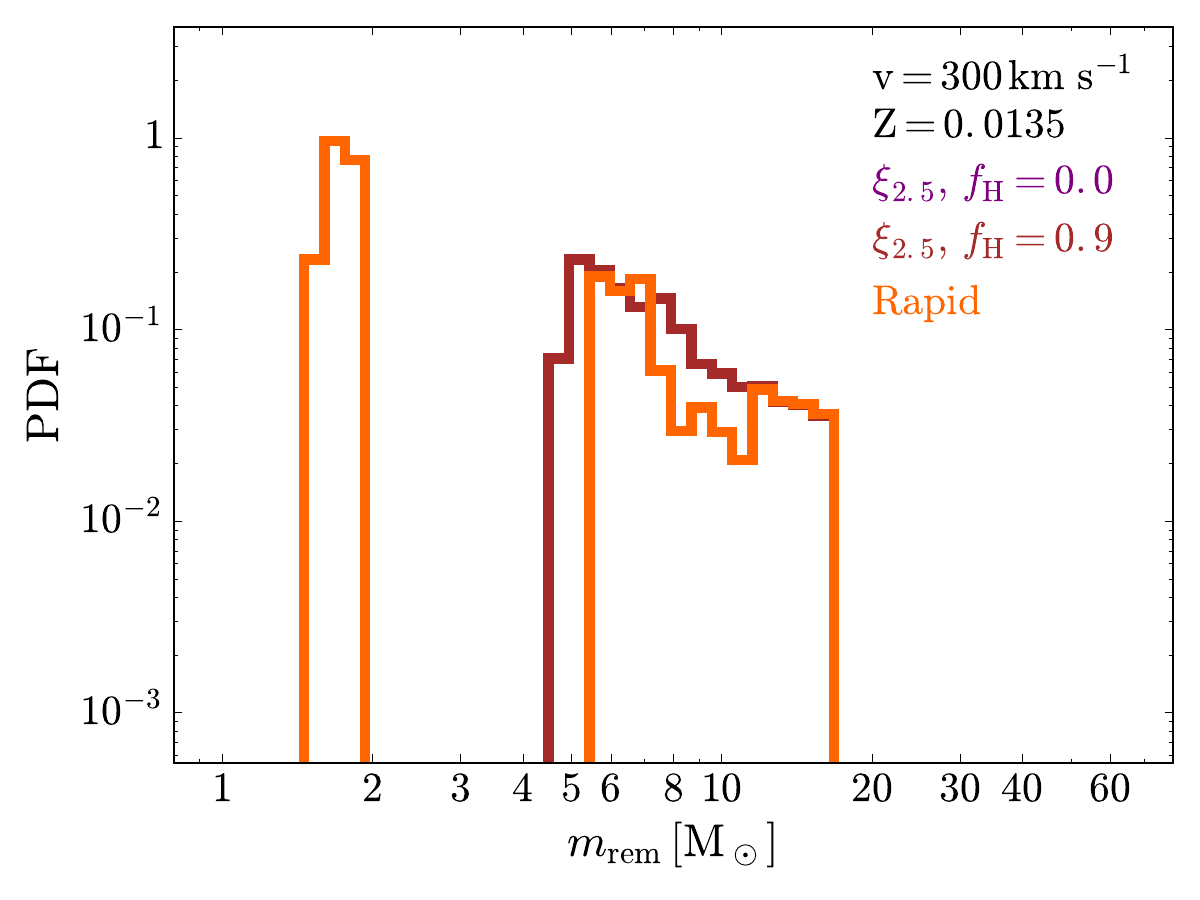}{0.33\textwidth}{}
          }
\caption{Probability distribution function (PDF) of compact object masses. We assume a Kroupa (2001) IMF for the progenitor stars with minimum mass $m_{\rm min}=13$ M$_\odot$ and maximum mass $m_{\rm max}=120$ M$_\odot$. Orange line: rapid model for CCSNe \citep{fryer2012}; purple line: compactness criterion with $\xi_{2.5}=0.3$ and $f_{\rm H}=0$; dark red line: compactness criterion with $\xi_{2.5}=0.3$ and $f_{\rm H}=0.9$. At $Z=0.0135$ the purple line is nearly invisible, because it  overlaps with the dark red line. Upper row: $v=0$~km~s$^{-1}$; middle row: $v=150$~km~s$^{-1}$; lower row: $v=300$~km~s$^{-1}$. Left-hand column: $Z=0.0003$; middle column: $Z=0.003$; right-hand column: $Z=0.0135$.
\label{fig:IMF}}
\end{figure*}
\subsection{Impact of CCSN model on BH masses}\label{sec:ccSN}
Figure~\ref{fig:rapid} shows the mass of compact objects we obtain assuming the rapid CCSN model described in \cite{fryer2012}. 
In contrast, Figure~\ref{fig:compac} is based on the compactness criterion. By considering these different models, we want to quantify the uncertainty on BH mass deriving from  CCSN prescriptions.

The main sources of uncertainty are the amount of fallback, the minimum value of the compactness (or carbon-oxygen mass) required for direct collapse and the fate of the hydrogen envelope (if any). The rapid model by \cite{fryer2012} assumes that fallback can be efficient (mass accreted by fallback $m_{\rm fb}\ge{}0.2$ M$_\odot$) and that stars with carbon-oxygen core mass $m_{\rm CO}\ge{}11$ M$_\odot$ collapse to BH directly, including their hydrogen envelope (if any). In contrast, in the compactness model we assume no fallback at all and we require that stars with compactness $\xi{}_{2.5}\ge{}0.3$ collapse to BH directly. In the case of direct collapse with the compactness criterion, if $f_{\rm H}=0.0$ ($f_{\rm H}=0.9$) we assume that the hydrogen envelope does not collapse (90~\% of the hydrogen envelope collapses) to BH.

The main difference between the rapid model and the compactness model, which manifests regardless of stellar rotation and metallicity, is the minimum ZAMS mass to form a BH (Table~\ref{tab:table1}). This difference arises mostly from the adopted threshold for direct collapse. In fact, direct collapse happens in the rapid model if $m_{\rm CO}\ge{}11$ M$_\odot$, which (according to equation~\ref{eq:fitcomp}) corresponds to compactness threshold $\xi{}_{2.5}\ge{}0.45$. By increasing the threshold for direct collapse from $\xi{}_{2.5}=0.3$ to $\xi{}_{2.5}=0.45$, the compactness models produce approximately the same minimum ZAMS mass for BH formation as the rapid model.


Another feature of the rapid model which does not show up in the compactness-based models, regardless of stellar metallicity and rotation, is the complex behavior of BH mass for $m_{\rm ZAMS}\lesssim{}40$ M$_\odot$. This is a consequence of the sophisticated fitting formulas for fallback derived from \cite{fryer2012}.

If $m_{\rm ZAMS}\gtrsim{}40$ M$_\odot$, metallicity and rotation matter, as we have seen in the previous section. If stellar metallicity is high ($Z=0.0135$) and $m_{\rm ZAMS}\gtrsim{}40$ M$_\odot$, the mass of BHs in the rapid model and in the compactness models have a remarkably similar behavior. The reason is that stellar winds are very efficient in massive stars with $Z=0.0135$ (almost independently of rotation) and remove the entire envelope, leveling the differences among the considered models.

In contrast, if stellar metallicity is low ($Z\le{}0.003$) and $m_{\rm ZAMS}\gtrsim{}40$ M$_\odot$, the initial rotation becomes the crucial ingredient. If the star rotates, the minimum ZAMS mass for PPISN and PISN decreases significantly (see the previous section) and stellar winds are efficient even at low metallicity. The combination of these two effects removes the hydrogen envelope and even a fraction of the He core. For this reason, the rapid model and the two compactness models are indistinguishable for rotating stars with $Z\le{}0.003$ and  $m_{\rm ZAMS}\gtrsim{}40$ M$_\odot$.

If the star does not rotate, the BH mass for $40\lesssim{}m_{\rm ZAMS}/M_\odot{}\lesssim{}80$ and $Z\le{}0.0003$ dramatically depends on the collapse of the H envelope, because the star retains a large portion of its hydrogen envelope to the final stages. Models assuming that most of the H envelope collapses (i.e. the rapid model and the compactness model with $f_{\rm H}=0.9$) predict a BH mass $m_{\rm BH}\sim{}60$ M$_\odot$,  almost twice as large as that expected from the compactness model with $f_{\rm H}=0$ in this range of ZAMS masses. Finally, non rotating stars with $m_{\rm ZAMS}\gtrsim{}80$ M$_\odot{}$ eject their H envelope entirely. Thus, the three CCSN models predict similar BH masses for extremely massive metal-poor non rotating stars.

In summary, if we look at the maximum BH mass, rotating models predict $m_{\rm BH,\,{}max}\leq{}45$ M$_\odot$ (originating from stars with $m_{\rm ZAMS}\sim{}90-100$ M$_\odot$ and $Z\leq{}0.0003$), regardless of the CCSN model.  In contrast, non-rotating models predict $m_{\rm BH,\,{}max}\sim{}60$ M$_\odot$ (originating from stars with $m_{\rm ZAMS}\sim{}60-70$ M$_\odot$ and $Z\leq{}0.0003$) if the H envelope is assumed to collapse, and  $m_{\rm BH,\,{}max}\sim{}45-50$ M$_\odot$ (originating from stars with  $m_{\rm ZAMS}\sim{}110-120$ M$_\odot$ and $Z\leq{}0.0003$) if the H envelope is assumed to be ejected. These conclusions depend on the adopted description of PPISNe and PISNe (from \citealt{spera2017}).


\subsection{Impact of rotation, CCSN model and metallicity on BH mass function}\label{sec:IMF}
For each considered metallicity, 
for each rotation speed and for each CCSN model separately, we have generated a set of $10^5$ single stars distributed according to a Kroupa initial mass function (IMF, i.e. $dN/dm\propto{}m^{-\alpha{}}$ with $\alpha=2.3$, \citealt{kroupa2001}), with minimum ZAMS mass $m_{\rm min}=13$ M$_\odot$ and maximum ZAMS mass $m_{\rm max}=120$ M$_\odot$.

Figure~\ref{fig:IMF} shows the mass function of compact objects for the three considered rotation speeds, for three metallicities ($Z=0.00003$ is not shown because it is almost indistinguishable from $Z=0.0003$) and for the three CCSN models. Note that the NS population is severely incomplete, because the minimum ZAMS mass currently available in the {\sc franec} tracks is $m_{\rm ZAMS}=13$ M$_\odot$. Smaller masses will be included in follow-up works. 

In general, the mass function of single BHs can be approximated with a power law, but the slope of the power law  depends on metallicity, on rotation speed and on the assumed CCSN prescription. 
If we make a linear fit of $\log_{10}{\rm PDF}=\mathcal{D}\,{}\log{}_{10} m_{\rm rem}+\mathcal{G}$ across our models, we find a preferred value of $\mathcal{D}\approx{}-0.5$, with a very large scatter. Binary evolution can change this scaling dramatically and will be included in a follow-up study.

The main differences among all the considered models 
are the number of NSs and the minimum mass of BHs. Because of the difference in the minimum ZAMS mass to form a BH (see Sections~\ref{sec:rotation} and \ref{sec:ccSN}), stars with $v=300$~km~s$^{-1}$ and minimum mass $m_{\rm ZAMS}=13$ M$_\odot$ adopting a compactness-based CCSN criterion do not form NSs, regardless of their metallicity. For these extremely fast rotating models to produce NSs, we need to assume a significantly higher $\xi{}_{2.5}$ threshold.

The minimum BH mass spans from $\sim{}4.5$ M$_\odot$ to $\sim{}15$ M$_\odot$, depending on the CCSN prescription (the compactness-based model with $f_{\rm H}=0.9$ produces significantly larger minimum BH masses at low metallicity) and on metallicity (metal-rich populations tend to produce BHs with a smaller minimum BH mass). The maximum BH mass dramatically depends  not only on metallicity, but also on rotation (BHs with mass $m_{\rm rem}\gtrsim{}60$~M$_\odot$ form only from non-rotating models).

At solar metallicity, the three CCSNe models and the three rotation speeds produce very similar BH populations (almost identical in the case of the two compactness-based models). The reason is that stellar winds peel-off massive stars, regardless of their initial rotation velocity and of the assumed CCSN model. In contrast, at lower metallicities the differences between the three CCSN models become important.

In this section we assumed that stars in the same stellar population have the same initial rotation speed. This is clearly a simplistic assumption because stars might form with different initial speed. Data of stellar rotation in the Milky Way show that stellar speeds should be distributed according to a Gaussian with average speed $\sim{}200$ km s$^{-1}$ and dispersion $\sim{}100$ km s$^{-1}$ \citep{duftonetal06}. In follow-up studies we will consider a distribution of initial stellar rotation velocities.


\begin{figure*}
  \gridline{\fig{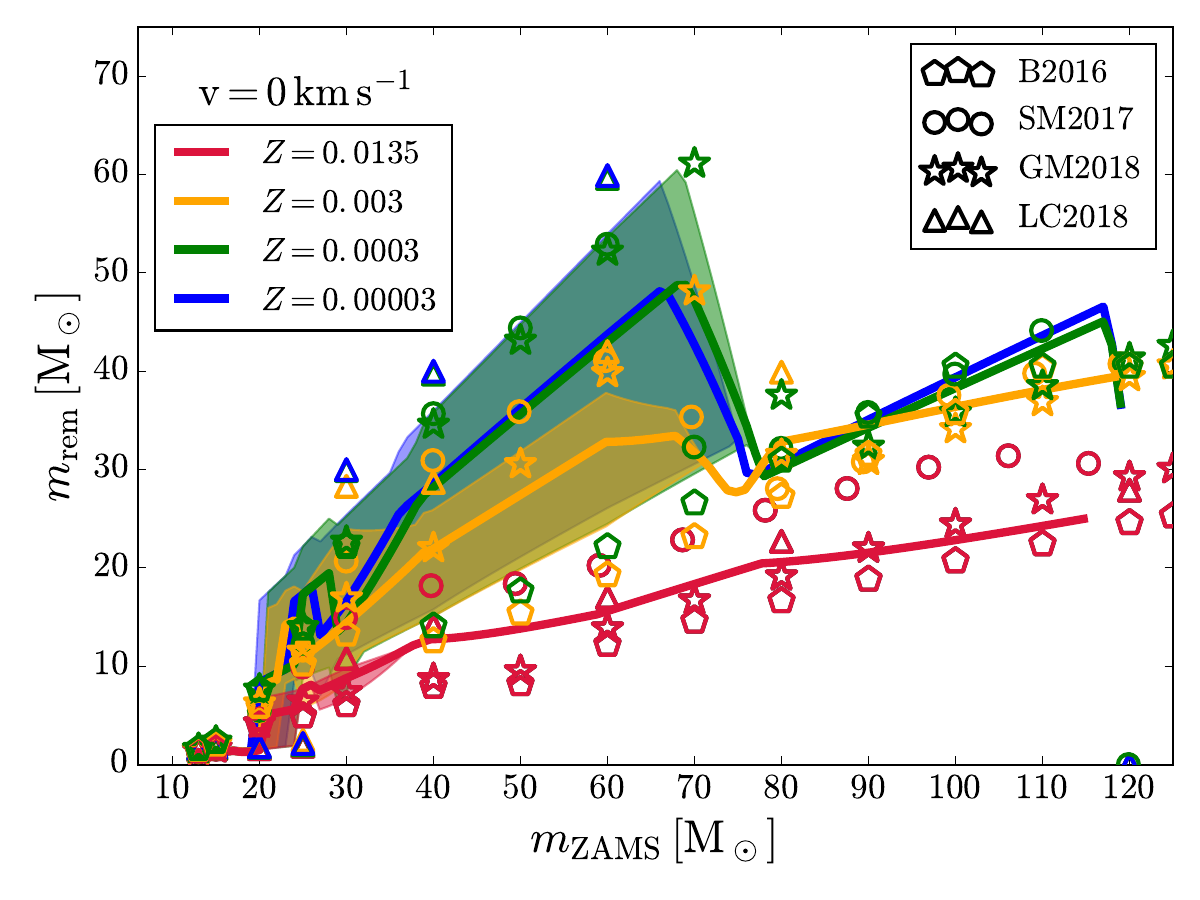}{0.33\textwidth}{}
    \fig{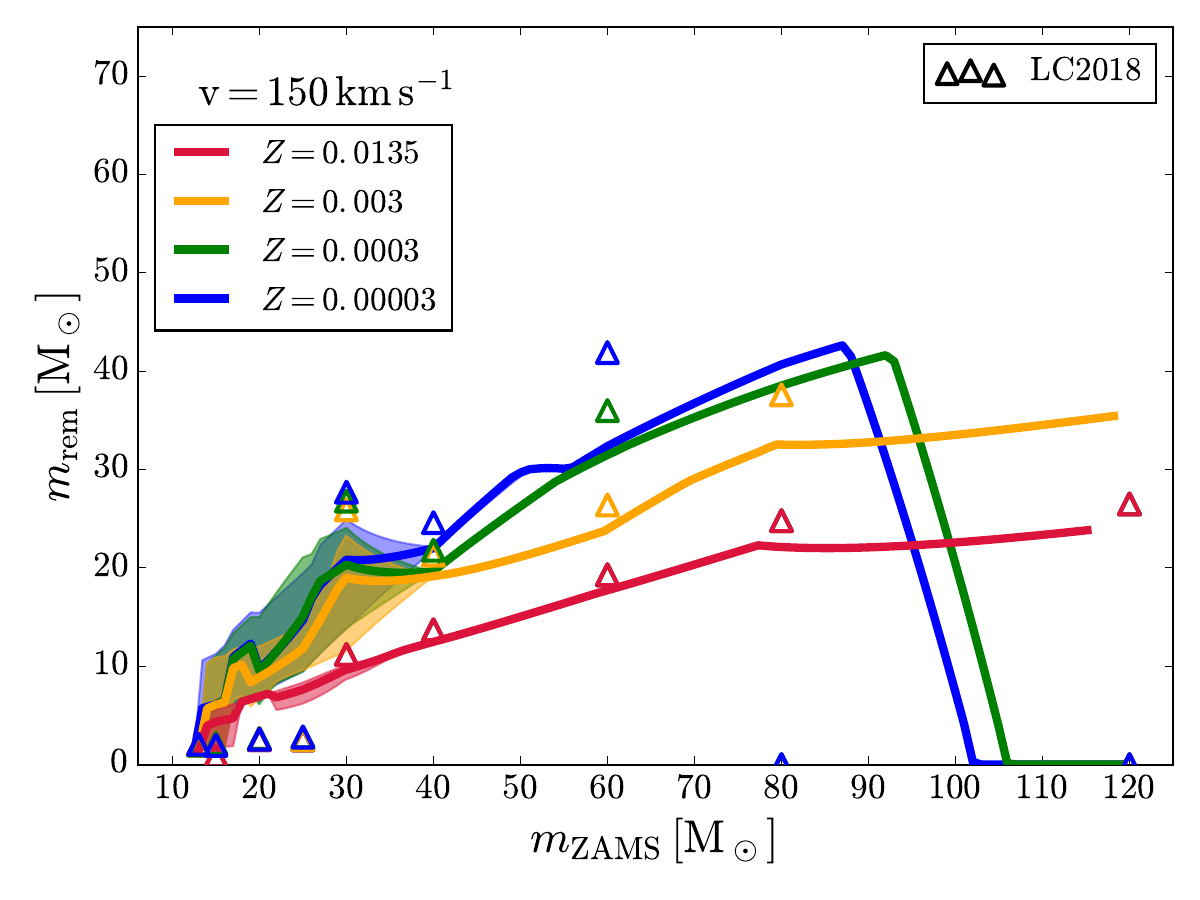}{0.33\textwidth}{}
    \fig{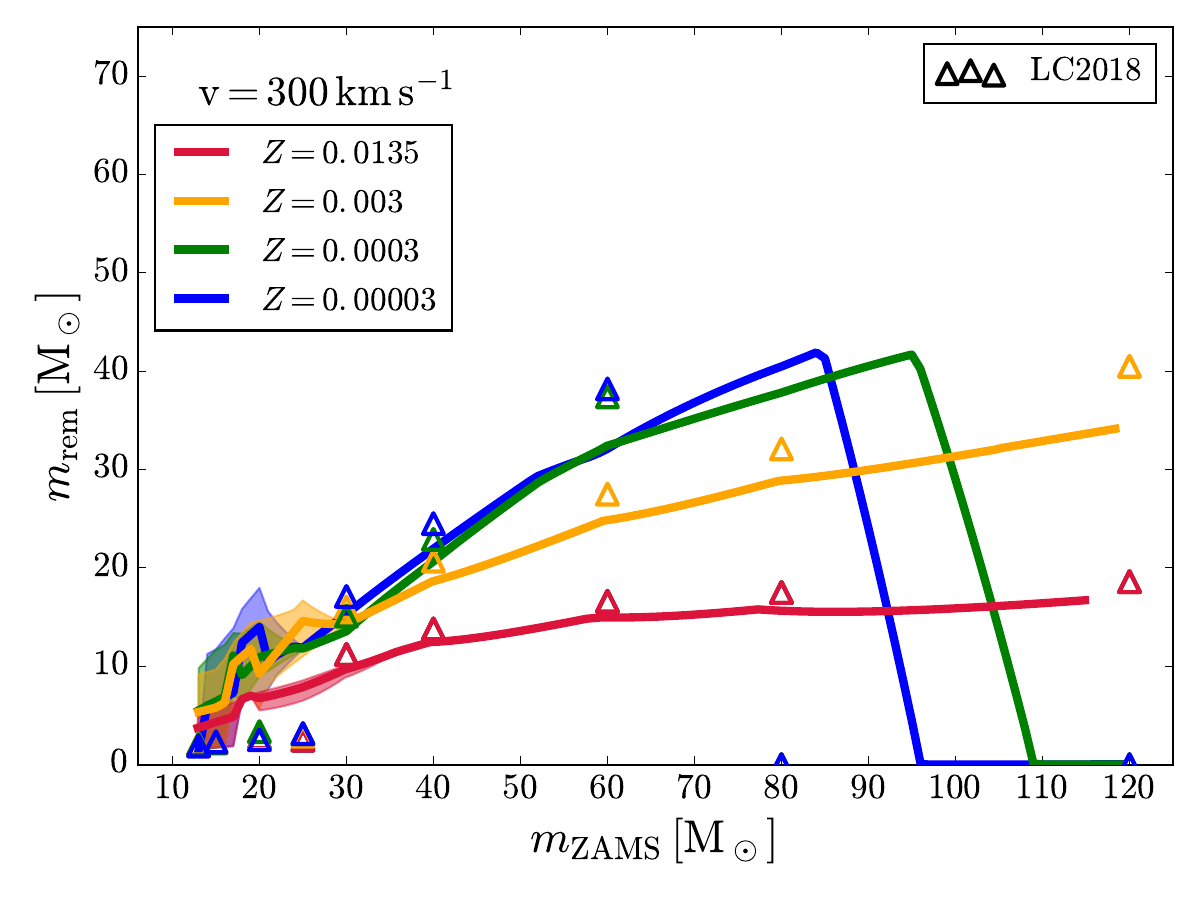}{0.33\textwidth}{}}
\caption{Mass of the compact object as a function of the progenitor's ZAMS mass for the main models we have considered, in comparison with some previous studies. Left: rotation velocity $v=0$ km s$^{-1}$. Middle: $v=150$ km s$^{-1}$. Right:  $v=300$ km s$^{-1}$.  The solid lines show the mean value of $m_{\rm rem}$ we obtain by averaging over the three CCSN models considered in this study, while the shaded areas show the maximum differences between the three CCSN models.
  Open triangles (LC2018): R model from \cite{limongi2018}. Open stars (GM2018): compact object mass predicted by  {\sc mobse} \citep{giacobbo2018b}, adopting the delayed CCSN model by \cite{fryer2012}. Open circles (SM2017): compact object mass estimated with {\sc sevn} \citep{spera2017}, adopting the delayed CCSN and the {\sc parsec} stellar tracks \citep{bressan2012}. Open pentagons (B2016): compact object mass estimated with {\sc bse} \citep{hurley2002}, adopting the same stellar winds, PPISN and PPISN model as {\sc startrack} \citep{belczynski2016pair}.
    In all panels and for all symbols and lines, red: progenitor's metallicity $Z=0.0135$; orange: $Z=0.003$; green: $Z=0.0003$; blue: $Z=0.00003$.\label{fig:comparison}}
\end{figure*}
\subsection{Comparison with previous work}
Figures~\ref{fig:rapid} and \ref{fig:compac} show that there is not much difference between {\sc parsec} models and {\sc franec} models with $v=0$~km~s$^{-1}$ when implemented inside {\sc sevn} and treated with the same model for CCSNe, PISNe and PPISNe. It is worth noting that while the typical difference in the maximum BH mass between {\sc franec} and {\sc parsec} is $\sim{}10$~\% at low metallicity,  the difference becomes $\sim{}27$~\% at solar metallicity ($Z=0.0135$, see Table~\ref{tab:table1}). This is explained with a different treatment of mass loss and different assumptions for chemical abundances.

Figure~\ref{fig:comparison} compares the mass spectrum of compact objects we derived in this study (considering only {\sc franec} tracks and accounting for the uncertainties induced by the CCSN model with a shaded area) with the mass spectrum obtained in previous studies, as a function of the ZAMS mass. In particular, we plot the mass spectrum from \cite{spera2017}, hereafter SM2017, from \cite{giacobbo2018b}, hereafter GM2018, and from \cite{limongi2018}, hereafter LC2018.  We also consider a version of {\sc bse} \citep{hurley2000,hurley2002} that includes the same stellar-wind, PISN and PPISN prescriptions as {\sc startrack} \citep{belczynski2016pair}, hereafter B2016. 


Our results are similar to the mass spectrum obtained with {\sc mobse} (GM2018), although the maximum BH mass in {\sc mobse} ($m_{\rm BH,\,{}max}\sim{}65$ M$_\odot$ at $Z=0.0003$) is $\sim{}8$~\% higher than the maximum mass we obtain with {\sc sevn}. Metal-poor stars with $m_{\rm ZAMS}\sim{}40-80$ M$_\odot$ seem to retain a more generous portion of their hydrogen envelope at collapse when integrated with {\sc mobse}. Our  results are also broadly consistent with SM2017 for metal-poor progenitors, while at $Z=0.0135$ SM2017 predict $\sim{}20-30$~\% larger BH masses (up to $\sim{}33$ M$_\odot$), explained by the fact that SM2017 adopt {\sc parsec} tracks. The models labelled as B2016 predict a maximum BH mass $\sim{}40$ M$_\odot$, significantly smaller than our model with $f_{\rm H}=0.9$ and similar to our model with H envelope ejection ($f_{\rm H}=0$). However, B2016 assume that the H envelope, when present, collapses with the rest of the star. In their model, metal-poor stars with $m_{\rm ZAMS}\sim{}40-80$ M$_\odot$ lose their hydrogen envelope almost completely for the different treatment of luminous blue variable stellar winds and of pulsational pair instability.

 LC2018 adopt the same {\sc franec} tracks we use here. Figure~\ref{fig:comparison} shows their model R which assumes that stars with $m_{\rm ZAMS}\le{}25$ M$_\odot$ explode as CCSNe, while stars with $m_{\rm ZAMS}>25$ M$_\odot$ collapse to BH directly, with $m_{\rm rem}=m_{\rm fin}$ (no mass ejection). Thus, the triangles shown in Figure~\ref{fig:comparison} represent the upper limit to BH masses we can obtain with {\sc franec} if $m_{\rm ZAMS}>25$~M$_\odot$.


 Finally, several previous studies  investigate the impact of stellar rotation on PPISNe and PISNe \citep{chatzopoulos2012a,chatzopoulos2012b,yoon2012,yusof2013,takahashi2018,uchida2019}. Our main findings agree with their results: i) the minimum ZAMS mass to undergo a PPISN and a PISN lowers significantly if stellar rotation is accounted for \citep{chatzopoulos2012a}, and ii) rotating models of massive stars lose their entire hydrogen-rich envelopes by enhanced mass loss \citep{yusof2013}.

\section{Conclusions}
We have investigated the impact of rotation and compactness on the mass of black holes (BHs), by implementing rotating stellar evolution models \citep{limongi2018} into our population synthesis code {\sc sevn} \citep{spera2015,spera2017,spera2019}.

Rotation has two major effects on BH formation. First, rotation reduces the minimum ZAMS mass for a star to collapse into a BH from $\sim{}18-25$ M$_\odot$ to $\sim{}13-18$ M$_\odot$ (according to the assumed CCSN prescriptions), because intermediate-mass ($m_{\rm ZAMS}\sim{}13-20$ M$_\odot$) rotating stars develop a larger carbon-oxygen core and a higher compactness than non-rotating stars.

Secondly, rotation reduces the maximum BH mass from metal-poor progenitors. This result comes from two combined effects: i) rotation increases stellar wind efficiency; thus, rotating metal-poor ($Z=0.00003-0.0003$) stars with $m_{\rm ZAMS}\sim{}40-80$ M$_\odot$ lose their H envelope entirely, while non-rotating metal-poor stars preserve most of it; ii) chemical mixing induced by rotation increases the mass of the He core, reducing the minimum ZAMS mass for PPISNe and PISNe to happen.

If we assume that the entire final mass of a star (including its residual hydrogen envelope) can collapse to a BH directly, the maximum BH mass from non-rotating stars is $\sim{}60$ M$_\odot$, while the maximum BH mass from fast rotating stars is $\sim{}45$ M$_\odot$.



Besides rotation, the mass of BHs is also strongly affected by the assumed CCSN model, especially by the amount of fallback, by the adopted threshold for direct collapse (based on $\xi{}_{2.5}$ or on $m_{\rm CO}$) and by the different fraction of hydrogen envelope that is able to collapse ($f_{\rm H}$).

In particular, the minimum ZAMS mass for a star to form a BH depends on the assumed threshold of compactness $\xi_{2.5}$ (larger values of the threshold  leading to higher minimum ZAMS masses) and on the efficiency of fallback.

The maximum BH mass that we expect from non rotating metal-poor ($Z=0.00003-0.0003$) stars depends wildly on the assumed CCSN prescription: if we assume that the residual hydrogen envelope participates in the collapse, the  maximum  BH mass is up to $\sim{}60$ M$_\odot$, approximately 1.5 times higher than if we assume that only the He core is able to collapse. This assumption is not important for metal-poor massive rotating stars and for metal-rich (both rotating and non-rotating) stars, because stellar winds remove their hydrogen envelope entirely, leveling these differences.




Here, we consider only single stars. In future works, we will investigate how binary evolution and star cluster dynamics affect our conclusions. We anticipate that close binary evolution should lead to a further stripping of the hydrogen envelope, affecting the maximum BH mass (see e.g. \citealt{giacobbo2018b}). On the other hand, star cluster dynamics can lead to the formation of binary BHs that incorporate the most massive BHs formed from single star evolution and from the merger of massive binaries (see e.g. \citealt{mapelli2016,dicarlo2019}), making the final scenario even more complex. 

 The methodology we presented here might be applied to estimate upper limits on BH spins. For all our models we find an upper limit to the final spin close to maximally rotating BHs. However, our models do not include mechanisms for efficient angular momentum dissipation, such as the Tayler-Spruit dynamo \citep{spruit2002,fuller2019}. Efficient angular momentum transport can lead to significantly lower BH spins ($a_{\rm BH}\lesssim{}0.1$), as described in several works (e.g. \citealt{heger2005,belczynski2017spin,qin2019,fuller2019b}). In a follow-up study, we will apply different models of angular momentum transport to our methodology.

Overall, we confirm that both stellar rotation and supernova prescriptions have a crucial impact on the mass function of BHs. This result provides a key to interpret future gravitational-wave data and to constrain stellar evolution and CCSN mechanisms.

\begin{deluxetable*}{cccccccc}
\tablenum{1}
\tablecaption{Most relevant masses.\label{tab:table1}}
\tablewidth{0pt}
\tablehead{
  \colhead{Stellar Ev.} & \colhead{CCSN} & \colhead{$v$ (km s$^{-1}$)} & $Z$  & \colhead{$m_{\rm ZAMS,\,{}min}$ (M$_\odot$)} & \colhead{$m_{\rm PPISN}$ (M$_\odot$)} & \colhead{$m_{\rm PISN}$ (M$_\odot$)} & \colhead{$m_{\rm BH,\,{}max}$ (M$_\odot$)}
}
\startdata
    {\sc franec} & rapid & 300 & 0.00003 & 18 & 52 & 97  & 42 \\
    {\sc franec} & rapid & 300 & 0.0003  & 17 & 53 & 110 & 42 \\
    {\sc franec} & rapid & 300 & 0.003   & 17 & 80 & --  & 34\\
    {\sc franec} & rapid & 300 & 0.0135  & 18 & -- & --  & 16     \\ 
    {\sc franec} & rapid & 150 & 0.00003 & 17 & 50 & 103 &43 \\
    {\sc franec} & rapid & 150 & 0.0003  & 17 & 55 & 107 & 42 \\
    {\sc franec} & rapid & 150 & 0.003   & 17 & 70 & --  & 35\\
    {\sc franec} & rapid & 150 & 0.0135  & 18 & -- & --  & 23    \\
    {\sc franec} & rapid &   0 & 0.00003 & 24 & 67 & -- &59 \\
    {\sc franec} & rapid &   0 & 0.0003  & 25 & 69 & -- & 60 \\
    {\sc franec} & rapid &   0 & 0.003   & 23 & 68 & -- & 39\\
    {\sc franec} & rapid &   0 & 0.0135  & 25 & -- & -- & 24     \\
    {\sc parsec} & rapid &   0 & 0.0003  & 22 & 63 & -- & 54 \\
    {\sc parsec} & rapid &   0 & 0.003   & 22 & 66 & -- & 43\\
    {\sc parsec} & rapid &   0 & 0.0135  & 23 & 94 & -- & 33   \\
    {\sc franec} & $\xi{}_{2.5}=0.3,\,{}f_{\rm H}=0$ & 300 & 0.00003 & 14 & 52 & 97  &42 \\
    {\sc franec} & $\xi{}_{2.5}=0.3,\,{}f_{\rm H}=0$ & 300 & 0.0003  & $\leq{}13$ & 53 & 110 &42 \\
    {\sc franec} & $\xi{}_{2.5}=0.3,\,{}f_{\rm H}=0$ & 300 & 0.003   & $\leq{}13$ & 80 & --  & 34\\
    {\sc franec} & $\xi{}_{2.5}=0.3,\,{}f_{\rm H}=0$ & 300 & 0.0135  & $\leq{}13$ & -- & --  &16 \\
    {\sc franec} & $\xi{}_{2.5}=0.3,\,{}f_{\rm H}=0$ & 150 & 0.00003 & 13 & 50 & 103 &43 \\
    {\sc franec} & $\xi{}_{2.5}=0.3,\,{}f_{\rm H}=0$ & 150 & 0.0003  & 14 & 55 & 107 & 42 \\
    {\sc franec} & $\xi{}_{2.5}=0.3,\,{}f_{\rm H}=0$ & 150 & 0.003   & 14 & 70 & -- & 35\\
    {\sc franec} & $\xi{}_{2.5}=0.3,\,{}f_{\rm H}=0$ & 150 & 0.0135  & 14 & -- & -- &23 \\
    {\sc franec} & $\xi{}_{2.5}=0.3,\,{}f_{\rm H}=0$ &   0 & 0.00003 & 20 & 67 & -- &47 \\
    {\sc franec} & $\xi{}_{2.5}=0.3,\,{}f_{\rm H}=0$ &   0 & 0.0003  & 21 & 69 & -- & 45 \\
    {\sc franec} & $\xi{}_{2.5}=0.3,\,{}f_{\rm H}=0$ &   0 & 0.003   & 21 & 68 & -- & 39\\
    {\sc franec} & $\xi{}_{2.5}=0.3,\,{}f_{\rm H}=0$ &   0 & 0.0135  & 21 & -- & -- &24 \\
    {\sc parsec} & $\xi{}_{2.5}=0.3,\,{}f_{\rm H}=0$ &   0 & 0.0003  & 19 & 63 & -- & 45 \\
    {\sc parsec} & $\xi{}_{2.5}=0.3,\,{}f_{\rm H}=0$ &   0 & 0.003   & 18 & 66 & -- & 41\\
    {\sc parsec} & $\xi{}_{2.5}=0.3,\,{}f_{\rm H}=0$ &   0 & 0.0135  & 19 & 94 & -- & 33 \\
    {\sc franec} & $\xi{}_{2.5}=0.3,\,{}f_{\rm H}=0.9$ & 300 & 0.00003 & 14 & 52 & 97 &42 \\
    {\sc franec} & $\xi{}_{2.5}=0.3,\,{}f_{\rm H}=0.9$ & 300 & 0.0003  & $\leq{}13$ & 53 & 110 &42 \\
    {\sc franec} & $\xi{}_{2.5}=0.3,\,{}f_{\rm H}=0.9$ & 300 & 0.003   & $\leq{}13$ &  80 & -- & 34\\
    {\sc franec} & $\xi{}_{2.5}=0.3,\,{}f_{\rm H}=0.9$ & 300 & 0.0135  & $\leq{}13$ &  -- & -- &16 \\
    {\sc franec} & $\xi{}_{2.5}=0.3,\,{}f_{\rm H}=0.9$ & 150 & 0.00003 & 13 & 50 & 103 &43 \\
    {\sc franec} & $\xi{}_{2.5}=0.3,\,{}f_{\rm H}=0.9$ & 150 & 0.0003  & 14 & 55 & 107 & 42 \\
    {\sc franec} & $\xi{}_{2.5}=0.3,\,{}f_{\rm H}=0.9$ & 150 & 0.003   & 14 & 70 & -- & 35\\
    {\sc franec} & $\xi{}_{2.5}=0.3,\,{}f_{\rm H}=0.9$ & 150 & 0.0135  & 14 & -- & -- &23    \\
    {\sc franec} & $\xi{}_{2.5}=0.3,\,{}f_{\rm H}=0.9$ &   0 & 0.00003 & 20 & 67 & -- &57 \\
    {\sc franec} & $\xi{}_{2.5}=0.3,\,{}f_{\rm H}=0.9$ &   0 & 0.0003  & 21 & 69 & -- & 58 \\
    {\sc franec} & $\xi{}_{2.5}=0.3,\,{}f_{\rm H}=0.9$ &   0 & 0.003   & 21 & 68 & -- & 39\\
    {\sc franec} & $\xi{}_{2.5}=0.3,\,{}f_{\rm H}=0.9$ &   0 & 0.0135  & 21 & -- & -- &24 \\    
    {\sc parsec} & $\xi{}_{2.5}=0.3,\,{}f_{\rm H}=0.9$ &  0 & 0.0003   & 19 & 63 & -- & 52 \\
    {\sc parsec} & $\xi{}_{2.5}=0.3,\,{}f_{\rm H}=0.9$ &  0 & 0.003    & 18 & 66 & -- & 42\\
    {\sc parsec} & $\xi{}_{2.5}=0.3,\,{}f_{\rm H}=0.9$ &  0 & 0.0135   & 19 & 94 & -- & 33 \\    
\enddata
\tablecomments{Column~(1): Stellar evolution tables (from {\sc franec} or {\sc parsec}). Column~(2): model for CCSN outcome (see Section~\ref{sec:SNe}). Column~(3): initial rotation speed of progenitor stars. Column~(4): progenitor's metallicity. Column~(5): minimum ZAMS mass to collapse to a BH (instead of producing a NS);  Column~(6): minimum ZAMS mass to undergo PPISN ($m_{\rm PPISN}$). Column~(7): minimum ZAMS mass to undergo PISN ($m_{\rm PISN}$). Column~(8): maximum BH mass ($m_{\rm BH,\,{}max}$).}
\end{deluxetable*}

\acknowledgments
We thank the anonymous referee for their careful reading of our manuscript and useful comments. MM acknowledges financial support by the European Research Council for the ERC Consolidator grant DEMOBLACK, under contract no. 770017.
MS acknowledges funding from the European Union's Horizon 2020 research and innovation programme under the Marie-Sklodowska-Curie grant agreement No. 794393.
 We acknowledge financial support from  MIUR through Progetto Premiale 'FIGARO'.


\software{{\sc sevn} \citep{spera2017},  
          {\sc parsec} \citep{bressan2012}, 
          {\sc franec} \citep{limongi2000},
          {\sc mobse} \citep{giacobbo2018}
          }



\appendix{}
\section{Fitting formula for PPISNe and PISNe}\label{sec:appendix}
When PISNe and PPISNe are effective, we derive the mass of the compact object as $m_{\rm rem} = \alpha_{\rm P}\,{}m_{\rm no\,{}PPI}$, where $m_{\rm no\,{}PPI}$ is the mass of the compact remnant we would obtain without PPISN/PISN. First, we define the following quantities
\begin{equation}
\mathcal{F} \equiv \dfrac{m_{\rm He}}{m_{\rm fin}},\quad{}
\mathcal{K} \equiv 0.67000\mathcal{F}+0.10000,\quad{}
\mathcal{S} \equiv 0.52260\mathcal{F}-0.52974.
\end{equation}
We then express $\alpha_{\mathrm{P}}$ as a function of $\mathcal{F}$, $\mathcal{S}$, $\mathcal{K}$ and $m_{\rm He}$:
\begin{equation}
\alpha{}_{\mathrm{P}} =
\begin{cases}
1\text{\hspace{15 pt} if } m_{\rm He}{}\leq 32{\rm M}_\odot{}, \,\, \forall \mathcal{F}, \,\, \forall \mathcal{S}\\

0.2\left(\mathcal{K}-1\right)m_{\rm He}{}+0.2\left(37-32\mathcal{K}\right)\,{}\text{\hspace{15 pt} if } 32<m_{\rm He}{}/{\rm M}_\odot{}\leq 37, \,\, \mathcal{F}<0.9, \,\, \forall \mathcal{S}\\

\mathcal{K}\,{}\text{\hspace{15 pt} if } 37<m_{\rm He}{}/{\rm M}_\odot{}\leq 60, \,\, \mathcal{F}<0.9, \,\, \forall \mathcal{S}\\

\mathcal{K}\left(16.0 - 0.25m_{\rm He}{}\right)\,{}\text{\hspace{15 pt} if } 60<m_{\rm He}{}/{\rm M}_\odot{}< 64, \,\, \mathcal{F}<0.9, \,\, \forall \mathcal{S}\\

\mathcal{S}\left(m_{\rm He}{}-32\right) + 1 \text{\hspace{15 pt} if } m_{\rm He}{}\leq 37{\rm M}_\odot{}, \,\, \mathcal{F}\geq 0.9, \,\, \forall \mathcal{S}\\

5\mathcal{S}+1 \text{\hspace{15 pt} if } 37<m_{\rm He}{}/{\rm M}_\odot{}\leq 56, \,\, \mathcal{F}\geq 0.9,\,\, 5\mathcal{S}+1<0.82916\\

\left(-0.1381\mathcal{F}+0.1309\right)\left(m_{\rm He}{}-56\right)+0.82916 \text{\hspace{15 pt} if } 37<m_{\rm He}{}/{\rm M}_\odot{}\leq 56, \,\, \mathcal{F}\geq 0.9,\,{}5\mathcal{S}+1\geq 0.82916\\
-0.103645m_{\rm He}{}+6.63328 \text{\hspace{15 pt} if } 56<m_{\rm He}{}/{\rm M}_\odot{}< 64, \,\, \mathcal{F}\geq 0.9, \,\, \forall \mathcal{S}\\

0 \text{\hspace{15 pt} if } 64\leq m_{\rm He}{}/{\rm M}_\odot{}< 135, \,\, \forall \mathcal{F},\,\, \forall \mathcal{S}\\

1 \text{\hspace{15 pt} if } m_{\rm He}{}\geq 135{\rm M}_\odot{}, \,\, \forall \mathcal{F}, \,\, \forall \mathcal{S}.\\
\end{cases}
\end{equation}
These fits are the same as we adopted in \cite{spera2017}, but here we fix some typos of Appendix~B of \cite{spera2017} (these typos did not affect the results of \citealt{spera2017}, because the code contained the correct equations).

\bibliography{bibliography}{}

\begin{thebibliography}{}
\expandafter\ifx\csname natexlab\endcsname\relax\def\natexlab#1{#1}\fi
\providecommand{\url}[1]{\href{#1}{#1}}
\providecommand{\dodoi}[1]{doi:~\href{http://doi.org/#1}{\nolinkurl{#1}}}
\providecommand{\doeprint}[1]{\href{http://ascl.net/#1}{\nolinkurl{http://ascl.net/#1}}}
\providecommand{\doarXiv}[1]{\href{https://arxiv.org/abs/#1}{\nolinkurl{https://arxiv.org/abs/#1}}}

\bibitem[{{Abbott}(2018{\natexlab{a}})}]{abbottO2}
{Abbott}, B. P. e.~a. 2018{\natexlab{a}}, ArXiv e-prints.
\newblock \doarXiv{1811.12907}

\bibitem[{{Abbott}(2018{\natexlab{b}})}]{abbottO2popandrate}
---. 2018{\natexlab{b}}, ArXiv e-prints.
\newblock \doarXiv{1811.12940}

\bibitem[{{Asplund} {et~al.}(2009){Asplund}, {Grevesse}, {Sauval}, \&
  {Scott}}]{agss09}
{Asplund}, M., {Grevesse}, N., {Sauval}, A.~J., \& {Scott}, P. 2009, \araa, 47,
  481, \dodoi{10.1146/annurev.astro.46.060407.145222}

\bibitem[{{Barkat} {et~al.}(1967){Barkat}, {Rakavy}, \& {Sack}}]{barkat1967}
{Barkat}, Z., {Rakavy}, G., \& {Sack}, N. 1967, Physical Review Letters, 18,
  379, \dodoi{10.1103/PhysRevLett.18.379}

\bibitem[{{Belczynski} {et~al.}(2010){Belczynski}, {Bulik}, {Fryer}, {Ruiter},
  {Valsecchi}, {Vink}, \& {Hurley}}]{belczynski2010}
{Belczynski}, K., {Bulik}, T., {Fryer}, C.~L., {et~al.} 2010, \apj, 714, 1217,
  \dodoi{10.1088/0004-637X/714/2/1217}

\bibitem[{{Belczynski} {et~al.}(2002){Belczynski}, {Kalogera}, \&
  {Bulik}}]{belczynski2002}
{Belczynski}, K., {Kalogera}, V., \& {Bulik}, T. 2002, \apj, 572, 407,
  \dodoi{10.1086/340304}

\bibitem[{{Belczynski} {et~al.}(2008){Belczynski}, {Kalogera}, {Rasio}, {Taam},
  {Zezas}, {Bulik}, {Maccarone}, \& {Ivanova}}]{belczynski2008}
{Belczynski}, K., {Kalogera}, V., {Rasio}, F.~A., {et~al.} 2008, \apjs, 174,
  223, \dodoi{10.1086/521026}

\bibitem[{{Belczynski} {et~al.}(2016){Belczynski}, {Heger}, {Gladysz},
  {Ruiter}, {Woosley}, {Wiktorowicz}, {Chen}, {Bulik}, {O'Shaughnessy}, {Holz},
  {Fryer}, \& {Berti}}]{belczynski2016pair}
{Belczynski}, K., {Heger}, A., {Gladysz}, W., {et~al.} 2016, \aap, 594, A97,
  \dodoi{10.1051/0004-6361/201628980}

\bibitem[{{Belczynski} {et~al.}(2017){Belczynski}, {Klencki}, {Fields},
  {Olejak}, {Berti}, {Meynet}, {Fryer}, {Holz}, {O'Shaughnessy}, {Brown},
  {Bulik}, {Leung}, {Nomoto}, {Madau}, {Hirschi}, {Jones}, {Mondal},
  {Chruslinska}, {Drozda}, {Gerosa}, {Doctor}, {Giersz}, {Ekstrom}, {Georgy},
  {Askar}, {Wysocki}, {Natan}, {Farr}, {Wiktorowicz}, {Miller}, {Farr}, \&
  {Lasota}}]{belczynski2017spin}
{Belczynski}, K., {Klencki}, J., {Fields}, C.~E., {et~al.} 2017, arXiv
  e-prints, arXiv:1706.07053.
\newblock \doarXiv{1706.07053}

\bibitem[{{Bethe} \& {Brown}(1998)}]{bethe1998}
{Bethe}, H.~A., \& {Brown}, G.~E. 1998, \apj, 506, 780, \dodoi{10.1086/306265}

\bibitem[{{Bressan} {et~al.}(2012){Bressan}, {Marigo}, {Girardi}, {Salasnich},
  {Dal Cero}, {Rubele}, \& {Nanni}}]{bressan2012}
{Bressan}, A., {Marigo}, P., {Girardi}, L., {et~al.} 2012, \mnras, 427, 127,
  \dodoi{10.1111/j.1365-2966.2012.21948.x}

\bibitem[{{Burrows} {et~al.}(2019){Burrows}, {Radice}, \&
  {Vartanyan}}]{burrows2019}
{Burrows}, A., {Radice}, D., \& {Vartanyan}, D. 2019, \mnras, 485, 3153,
  \dodoi{10.1093/mnras/stz543}

\bibitem[{{Burrows} {et~al.}(2018){Burrows}, {Vartanyan}, {Dolence}, {Skinner},
  \& {Radice}}]{burrows2018}
{Burrows}, A., {Vartanyan}, D., {Dolence}, J.~C., {Skinner}, M.~A., \&
  {Radice}, D. 2018, \ssr, 214, 33, \dodoi{10.1007/s11214-017-0450-9}

\bibitem[{{Cayrel} {et~al.}(2004){Cayrel}, {Depagne}, {Spite}, {Hill}, {Spite},
  {Fran{\c{c}}ois}, {Plez}, {Beers}, {Primas}, {Andersen}, {Barbuy},
  {Bonifacio}, {Molaro}, \& {Nordstr{\"o}m}}]{cayreletal04}
{Cayrel}, R., {Depagne}, E., {Spite}, M., {et~al.} 2004, \aap, 416, 1117,
  \dodoi{10.1051/0004-6361:20034074}

\bibitem[{{Chatzopoulos} \& {Wheeler}(2012{\natexlab{a}})}]{chatzopoulos2012a}
{Chatzopoulos}, E., \& {Wheeler}, J.~C. 2012{\natexlab{a}}, \apj, 748, 42,
  \dodoi{10.1088/0004-637X/748/1/42}

\bibitem[{{Chatzopoulos} \& {Wheeler}(2012{\natexlab{b}})}]{chatzopoulos2012b}
---. 2012{\natexlab{b}}, \apj, 760, 154, \dodoi{10.1088/0004-637X/760/2/154}

\bibitem[{{Chen} {et~al.}(2015){Chen}, {Bressan}, {Girardi}, {Marigo}, {Kong},
  \& {Lanza}}]{chen2015}
{Chen}, Y., {Bressan}, A., {Girardi}, L., {et~al.} 2015, \mnras, 452, 1068,
  \dodoi{10.1093/mnras/stv1281}

\bibitem[{{Chieffi} \& {Limongi}(2004)}]{chieffi2004}
{Chieffi}, A., \& {Limongi}, M. 2004, \apj, 608, 405, \dodoi{10.1086/392523}

\bibitem[{{Chieffi} \& {Limongi}(2013)}]{chieffi2013}
---. 2013, \apj, 764, 21, \dodoi{10.1088/0004-637X/764/1/21}

\bibitem[{{de Jager} {et~al.}(1988){de Jager}, {Nieuwenhuijzen}, \& {van der
  Hucht}}]{dejager88}
{de Jager}, C., {Nieuwenhuijzen}, H., \& {van der Hucht}, K.~A. 1988, \aaps,
  72, 259

\bibitem[{{de Mink} \& {Mandel}(2016)}]{demink2016}
{de Mink}, S.~E., \& {Mandel}, I. 2016, \mnras, 460, 3545,
  \dodoi{10.1093/mnras/stw1219}

\bibitem[{{Di Carlo} {et~al.}(2019){Di Carlo}, {Giacobbo}, {Mapelli},
  {Pasquato}, {Spera}, {Wang}, \& {Haardt}}]{dicarlo2019}
{Di Carlo}, U.~N., {Giacobbo}, N., {Mapelli}, M., {et~al.} 2019, \mnras, 487,
  2947, \dodoi{10.1093/mnras/stz1453}

\bibitem[{{Dufton} {et~al.}(2006){Dufton}, {Smartt}, {Lee}, {Ryans}, {Hunter},
  {Evans}, {Herrero}, {Trundle}, {Lennon}, {Irwin}, \& {Kaufer}}]{duftonetal06}
{Dufton}, P.~L., {Smartt}, S.~J., {Lee}, J.~K., {et~al.} 2006, \aap, 457, 265,
  \dodoi{10.1051/0004-6361:20065392}

\bibitem[{{Dvorkin} {et~al.}(2018){Dvorkin}, {Uzan}, {Vangioni}, \&
  {Silk}}]{dvorkin2018}
{Dvorkin}, I., {Uzan}, J.-P., {Vangioni}, E., \& {Silk}, J. 2018, \mnras, 479,
  121, \dodoi{10.1093/mnras/sty1414}

\bibitem[{{Ebinger} {et~al.}(2019{\natexlab{a}}){Ebinger}, {Curtis},
  {Fr{\"o}hlich}, {Hempel}, {Perego}, {Liebend{\"o}rfer}, \&
  {Thielemann}}]{ebinger2019a}
{Ebinger}, K., {Curtis}, S., {Fr{\"o}hlich}, C., {et~al.} 2019{\natexlab{a}},
  \apj, 870, 1, \dodoi{10.3847/1538-4357/aae7c9}

\bibitem[{{Ebinger} {et~al.}(2019{\natexlab{b}}){Ebinger}, {Curtis}, {Ghosh},
  {Fr{\"o}hlich}, {Hempel}, {Perego}, {Liebend{\"o}rfer}, \&
  {Thielemann}}]{ebinger2019b}
{Ebinger}, K., {Curtis}, S., {Ghosh}, S., {et~al.} 2019{\natexlab{b}}, arXiv
  e-prints, arXiv:1910.08958.
\newblock \doarXiv{1910.08958}

\bibitem[{{Ekstr{\"o}m} {et~al.}(2012){Ekstr{\"o}m}, {Georgy}, {Eggenberger},
  {Meynet}, {Mowlavi}, {Wyttenbach}, {Granada}, {Decressin}, {Hirschi},
  {Frischknecht}, {Charbonnel}, \& {Maeder}}]{ekstrom2012}
{Ekstr{\"o}m}, S., {Georgy}, C., {Eggenberger}, P., {et~al.} 2012, \aap, 537,
  A146, \dodoi{10.1051/0004-6361/201117751}

\bibitem[{{Eldridge} \& {Stanway}(2016)}]{eldridge2016}
{Eldridge}, J.~J., \& {Stanway}, E.~R. 2016, \mnras, 462, 3302,
  \dodoi{10.1093/mnras/stw1772}

\bibitem[{{Eldridge} {et~al.}(2019){Eldridge}, {Stanway}, \&
  {Tang}}]{eldridge2019}
{Eldridge}, J.~J., {Stanway}, E.~R., \& {Tang}, P.~N. 2019, \mnras, 482, 870,
  \dodoi{10.1093/mnras/sty2714}

\bibitem[{{Ertl} {et~al.}(2016){Ertl}, {Janka}, {Woosley}, {Sukhbold}, \&
  {Ugliano}}]{ertl2016}
{Ertl}, T., {Janka}, H.-T., {Woosley}, S.~E., {Sukhbold}, T., \& {Ugliano}, M.
  2016, \apj, 818, 124, \dodoi{10.3847/0004-637X/818/2/124}

\bibitem[{{Farr} {et~al.}(2011){Farr}, {Sravan}, {Cantrell}, {Kreidberg},
  {Bailyn}, {Mandel}, \& {Kalogera}}]{farr2011}
{Farr}, W.~M., {Sravan}, N., {Cantrell}, A., {et~al.} 2011, \apj, 741, 103,
  \dodoi{10.1088/0004-637X/741/2/103}

\bibitem[{{Fern{\'a}ndez} {et~al.}(2018){Fern{\'a}ndez}, {Quataert},
  {Kashiyama}, \& {Coughlin}}]{fernandez2018}
{Fern{\'a}ndez}, R., {Quataert}, E., {Kashiyama}, K., \& {Coughlin}, E.~R.
  2018, \mnras, 476, 2366, \dodoi{10.1093/mnras/sty306}

\bibitem[{{Foglizzo} {et~al.}(2015){Foglizzo}, {Kazeroni}, {Guilet}, {Masset},
  {Gonz{\'a}lez}, {Krueger}, {Novak}, {Oertel}, {Margueron}, {Faure}, {Martin},
  {Blottiau}, {Peres}, \& {Durand}}]{foglizzo2015}
{Foglizzo}, T., {Kazeroni}, R., {Guilet}, J., {et~al.} 2015, \pasa, 32, e009,
  \dodoi{10.1017/pasa.2015.9}

\bibitem[{{Fowler} \& {Hoyle}(1964)}]{fowler1964}
{Fowler}, W.~A., \& {Hoyle}, F. 1964, \apjs, 9, 201, \dodoi{10.1086/190103}

\bibitem[{{Fryer}(1999)}]{fryer1999}
{Fryer}, C.~L. 1999, \apj, 522, 413, \dodoi{10.1086/307647}

\bibitem[{{Fryer} {et~al.}(2012){Fryer}, {Belczynski}, {Wiktorowicz},
  {Dominik}, {Kalogera}, \& {Holz}}]{fryer2012}
{Fryer}, C.~L., {Belczynski}, K., {Wiktorowicz}, G., {et~al.} 2012, \apj, 749,
  91, \dodoi{10.1088/0004-637X/749/1/91}

\bibitem[{{Fryer} \& {Kalogera}(2001)}]{fryer2001}
{Fryer}, C.~L., \& {Kalogera}, V. 2001, \apj, 554, 548, \dodoi{10.1086/321359}

\bibitem[{{Fuller} \& {Ma}(2019)}]{fuller2019b}
{Fuller}, J., \& {Ma}, L. 2019, \apjl, 881, L1,
  \dodoi{10.3847/2041-8213/ab339b}

\bibitem[{{Fuller} {et~al.}(2019){Fuller}, {Piro}, \& {Jermyn}}]{fuller2019}
{Fuller}, J., {Piro}, A.~L., \& {Jermyn}, A.~S. 2019, \mnras, 485, 3661,
  \dodoi{10.1093/mnras/stz514}

\bibitem[{{Giacobbo} \& {Mapelli}(2018)}]{giacobbo2018b}
{Giacobbo}, N., \& {Mapelli}, M. 2018, \mnras, 480, 2011,
  \dodoi{10.1093/mnras/sty1999}

\bibitem[{{Giacobbo} {et~al.}(2018){Giacobbo}, {Mapelli}, \&
  {Spera}}]{giacobbo2018}
{Giacobbo}, N., {Mapelli}, M., \& {Spera}, M. 2018, \mnras, 474, 2959,
  \dodoi{10.1093/mnras/stx2933}

\bibitem[{{Gr{\"a}fener} \& {Hamann}(2008)}]{graefener2008}
{Gr{\"a}fener}, G., \& {Hamann}, W.-R. 2008, \aap, 482, 945,
  \dodoi{10.1051/0004-6361:20066176}

\bibitem[{{Groh} {et~al.}(2019){Groh}, {Ekstr{\"o}m}, {Georgy}, {Meynet},
  {Choplin}, {Eggenberger}, {Hirschi}, {Maeder}, {Murphy}, {Boian}, \&
  {Farrell}}]{groh2019}
{Groh}, J.~H., {Ekstr{\"o}m}, S., {Georgy}, C., {et~al.} 2019, \aap, 627, A24,
  \dodoi{10.1051/0004-6361/201833720}

\bibitem[{{Heger} {et~al.}(2003){Heger}, {Fryer}, {Woosley}, {Langer}, \&
  {Hartmann}}]{heger2003}
{Heger}, A., {Fryer}, C.~L., {Woosley}, S.~E., {Langer}, N., \& {Hartmann},
  D.~H. 2003, \apj, 591, 288, \dodoi{10.1086/375341}

\bibitem[{{Heger} {et~al.}(2000){Heger}, {Langer}, \& {Woosley}}]{hlw00}
{Heger}, A., {Langer}, N., \& {Woosley}, S.~E. 2000, \apj, 528, 368,
  \dodoi{10.1086/308158}

\bibitem[{{Heger} {et~al.}(2005){Heger}, {Woosley}, \& {Spruit}}]{heger2005}
{Heger}, A., {Woosley}, S.~E., \& {Spruit}, H.~C. 2005, \apj, 626, 350,
  \dodoi{10.1086/429868}

\bibitem[{{Horiuchi} {et~al.}(2014){Horiuchi}, {Nakamura}, {Takiwaki},
  {Kotake}, \& {Tanaka}}]{horiuchi2014}
{Horiuchi}, S., {Nakamura}, K., {Takiwaki}, T., {Kotake}, K., \& {Tanaka}, M.
  2014, \mnras, 445, L99, \dodoi{10.1093/mnrasl/slu146}

\bibitem[{{Hunter} {et~al.}(2008){Hunter}, {Lennon}, {Dufton}, {Trundle},
  {Sim{\'o}n-D{\'\i}az}, {Smartt}, {Ryans}, \& {Evans}}]{hunteretal08}
{Hunter}, I., {Lennon}, D.~J., {Dufton}, P.~L., {et~al.} 2008, \aap, 479, 541,
  \dodoi{10.1051/0004-6361:20078511}

\bibitem[{{Hunter} {et~al.}(2009){Hunter}, {Brott}, {Langer}, {Lennon},
  {Dufton}, {Howarth}, {Ryans}, {Trundle}, {Evans}, {de Koter}, \&
  {Smartt}}]{untetal09}
{Hunter}, I., {Brott}, I., {Langer}, N., {et~al.} 2009, \aap, 496, 841,
  \dodoi{10.1051/0004-6361/200809925}

\bibitem[{{Hurley} {et~al.}(2000){Hurley}, {Pols}, \& {Tout}}]{hurley2000}
{Hurley}, J.~R., {Pols}, O.~R., \& {Tout}, C.~A. 2000, \mnras, 315, 543,
  \dodoi{10.1046/j.1365-8711.2000.03426.x}

\bibitem[{{Hurley} {et~al.}(2002){Hurley}, {Tout}, \& {Pols}}]{hurley2002}
{Hurley}, J.~R., {Tout}, C.~A., \& {Pols}, O.~R. 2002, \mnras, 329, 897,
  \dodoi{10.1046/j.1365-8711.2002.05038.x}

\bibitem[{{Janka}(2012)}]{janka2012}
{Janka}, H.-T. 2012, Annual Review of Nuclear and Particle Science, 62, 407,
  \dodoi{10.1146/annurev-nucl-102711-094901}

\bibitem[{{Janka}(2017)}]{janka2017}
---. 2017, \apj, 837, 84, \dodoi{10.3847/1538-4357/aa618e}

\bibitem[{{Jones} {et~al.}(2013){Jones}, {Hirschi}, {Nomoto}, {Fischer},
  {Timmes}, {Herwig}, {Paxton}, {Toki}, {Suzuki}, {Mart{\'{\i}}nez-Pinedo},
  {Lam}, \& {Bertolli}}]{jones2013}
{Jones}, S., {Hirschi}, R., {Nomoto}, K., {et~al.} 2013, \apj, 772, 150,
  \dodoi{10.1088/0004-637X/772/2/150}

\bibitem[{{Kroupa}(2001)}]{kroupa2001}
{Kroupa}, P. 2001, \mnras, 322, 231, \dodoi{10.1046/j.1365-8711.2001.04022.x}

\bibitem[{{Kruckow} {et~al.}(2018){Kruckow}, {Tauris}, {Langer}, {Kramer}, \&
  {Izzard}}]{kruckow2018}
{Kruckow}, M.~U., {Tauris}, T.~M., {Langer}, N., {Kramer}, M., \& {Izzard},
  R.~G. 2018, \mnras, 481, 1908, \dodoi{10.1093/mnras/sty2190}

\bibitem[{{Kudritzki} {et~al.}(1987){Kudritzki}, {Pauldrach}, \&
  {Puls}}]{kudritzki1987}
{Kudritzki}, R.~P., {Pauldrach}, A., \& {Puls}, J. 1987, \aap, 173, 293

\bibitem[{{Limongi}(2017)}]{limongi2017}
{Limongi}, M. 2017, {Supernovae from Massive Stars}, 513,
  \dodoi{10.1007/978-3-319-21846-5_119}

\bibitem[{{Limongi} \& {Chieffi}(2006)}]{limongi2006}
{Limongi}, M., \& {Chieffi}, A. 2006, \apj, 647, 483, \dodoi{10.1086/505164}

\bibitem[{{Limongi} \& {Chieffi}(2018)}]{limongi2018}
---. 2018, \apjs, 237, 13, \dodoi{10.3847/1538-4365/aacb24}

\bibitem[{{Limongi} {et~al.}(2000){Limongi}, {Straniero}, \&
  {Chieffi}}]{limongi2000}
{Limongi}, M., {Straniero}, O., \& {Chieffi}, A. 2000, \apjs, 129, 625,
  \dodoi{10.1086/313424}

\bibitem[{{Lovegrove} \& {Woosley}(2013)}]{lovegrove2013}
{Lovegrove}, E., \& {Woosley}, S.~E. 2013, \apj, 769, 109,
  \dodoi{10.1088/0004-637X/769/2/109}

\bibitem[{{Mandel} \& {de Mink}(2016)}]{mandel2016}
{Mandel}, I., \& {de Mink}, S.~E. 2016, \mnras, 458, 2634,
  \dodoi{10.1093/mnras/stw379}

\bibitem[{{Mapelli}(2016)}]{mapelli2016}
{Mapelli}, M. 2016, \mnras, 459, 3432, \dodoi{10.1093/mnras/stw869}

\bibitem[{{Mapelli} {et~al.}(2009){Mapelli}, {Colpi}, \&
  {Zampieri}}]{mapelli2009}
{Mapelli}, M., {Colpi}, M., \& {Zampieri}, L. 2009, \mnras, 395, L71,
  \dodoi{10.1111/j.1745-3933.2009.00645.x}

\bibitem[{{Mapelli} \& {Giacobbo}(2018)}]{mapelli2018}
{Mapelli}, M., \& {Giacobbo}, N. 2018, \mnras, 479, 4391,
  \dodoi{10.1093/mnras/sty1613}

\bibitem[{{Mapelli} {et~al.}(2017){Mapelli}, {Giacobbo}, {Ripamonti}, \&
  {Spera}}]{mapelli2017}
{Mapelli}, M., {Giacobbo}, N., {Ripamonti}, E., \& {Spera}, M. 2017, \mnras,
  472, 2422, \dodoi{10.1093/mnras/stx2123}

\bibitem[{{Mapelli} {et~al.}(2019){Mapelli}, {Giacobbo}, {Santoliquido}, \&
  {Artale}}]{mapelli2019}
{Mapelli}, M., {Giacobbo}, N., {Santoliquido}, F., \& {Artale}, M.~C. 2019,
  \mnras, 487, 2, \dodoi{10.1093/mnras/stz1150}

\bibitem[{{Mapelli} {et~al.}(2010){Mapelli}, {Ripamonti}, {Zampieri}, {Colpi},
  \& {Bressan}}]{mapelli2010}
{Mapelli}, M., {Ripamonti}, E., {Zampieri}, L., {Colpi}, M., \& {Bressan}, A.
  2010, \mnras, 408, 234, \dodoi{10.1111/j.1365-2966.2010.17048.x}

\bibitem[{{Mapelli} {et~al.}(2013){Mapelli}, {Zampieri}, {Ripamonti}, \&
  {Bressan}}]{mapelli2013}
{Mapelli}, M., {Zampieri}, L., {Ripamonti}, E., \& {Bressan}, A. 2013, \mnras,
  429, 2298, \dodoi{10.1093/mnras/sts500}

\bibitem[{{Marchant} {et~al.}(2016){Marchant}, {Langer}, {Podsiadlowski},
  {Tauris}, \& {Moriya}}]{marchant2016}
{Marchant}, P., {Langer}, N., {Podsiadlowski}, P., {Tauris}, T.~M., \&
  {Moriya}, T.~J. 2016, \aap, 588, A50, \dodoi{10.1051/0004-6361/201628133}

\bibitem[{{Marigo} {et~al.}(2017){Marigo}, {Girardi}, {Bressan}, {Rosenfield},
  {Aringer}, {Chen}, {Dussin}, {Nanni}, {Pastorelli}, {Rodrigues}, {Trabucchi},
  {Bladh}, {Dalcanton}, {Groenewegen}, {Montalb{\'a}n}, \& {Wood}}]{marigo2017}
{Marigo}, P., {Girardi}, L., {Bressan}, A., {et~al.} 2017, \apj, 835, 77,
  \dodoi{10.3847/1538-4357/835/1/77}

\bibitem[{{Mennekens} \& {Vanbeveren}(2014)}]{mennekens2014}
{Mennekens}, N., \& {Vanbeveren}, D. 2014, \aap, 564, A134,
  \dodoi{10.1051/0004-6361/201322198}

\bibitem[{{Meynet} \& {Maeder}(2005)}]{meynet2005}
{Meynet}, G., \& {Maeder}, A. 2005, \aap, 429, 581,
  \dodoi{10.1051/0004-6361:20047106}

\bibitem[{{Nadezhin}(1980)}]{nadezhin1980}
{Nadezhin}, D.~K. 1980, \apss, 69, 115, \dodoi{10.1007/BF00638971}

\bibitem[{{Nomoto}(1984)}]{nomoto1984}
{Nomoto}, K. 1984, \apj, 277, 791, \dodoi{10.1086/161749}

\bibitem[{{Nomoto}(1987)}]{nomoto1987}
---. 1987, \apj, 322, 206, \dodoi{10.1086/165716}

\bibitem[{{Nugis} \& {Lamers}(2000)}]{nl00}
{Nugis}, T., \& {Lamers}, H.~J.~G.~L.~M. 2000, \aap, 360, 227

\bibitem[{{O'Connor} \& {Ott}(2011)}]{oconnor2011}
{O'Connor}, E., \& {Ott}, C.~D. 2011, \apj, 730, 70,
  \dodoi{10.1088/0004-637X/730/2/70}

\bibitem[{{{\"O}zel} \& {Freire}(2016)}]{oezel2016}
{{\"O}zel}, F., \& {Freire}, P. 2016, \araa, 54, 401,
  \dodoi{10.1146/annurev-astro-081915-023322}

\bibitem[{{{\"O}zel} {et~al.}(2010){{\"O}zel}, {Psaltis}, {Narayan}, \&
  {McClintock}}]{oezel2010}
{{\"O}zel}, F., {Psaltis}, D., {Narayan}, R., \& {McClintock}, J.~E. 2010,
  \apj, 725, 1918, \dodoi{10.1088/0004-637X/725/2/1918}

\bibitem[{{Paxton} {et~al.}(2011){Paxton}, {Bildsten}, {Dotter}, {Herwig},
  {Lesaffre}, \& {Timmes}}]{paxton2011}
{Paxton}, B., {Bildsten}, L., {Dotter}, A., {et~al.} 2011, \apjs, 192, 3,
  \dodoi{10.1088/0067-0049/192/1/3}

\bibitem[{{Paxton} {et~al.}(2013){Paxton}, {Cantiello}, {Arras}, {Bildsten},
  {Brown}, {Dotter}, {Mankovich}, {Montgomery}, {Stello}, {Timmes}, \&
  {Townsend}}]{paxton2013}
{Paxton}, B., {Cantiello}, M., {Arras}, P., {et~al.} 2013, \apjs, 208, 4,
  \dodoi{10.1088/0067-0049/208/1/4}

\bibitem[{{Paxton} {et~al.}(2015){Paxton}, {Marchant}, {Schwab}, {Bauer},
  {Bildsten}, {Cantiello}, {Dessart}, {Farmer}, {Hu}, {Langer}, {Townsend},
  {Townsley}, \& {Timmes}}]{paxton2015}
{Paxton}, B., {Marchant}, P., {Schwab}, J., {et~al.} 2015, \apjs, 220, 15,
  \dodoi{10.1088/0067-0049/220/1/15}

\bibitem[{{Pejcha} \& {Thompson}(2015)}]{pejcha2015}
{Pejcha}, O., \& {Thompson}, T.~A. 2015, \apj, 801, 90,
  \dodoi{10.1088/0004-637X/801/2/90}

\bibitem[{{Portegies Zwart} \& {Yungelson}(1998)}]{portegieszwart1998}
{Portegies Zwart}, S.~F., \& {Yungelson}, L.~R. 1998, \aap, 332, 173

\bibitem[{{Qin} {et~al.}(2019){Qin}, {Marchant}, {Fragos}, {Meynet}, \&
  {Kalogera}}]{qin2019}
{Qin}, Y., {Marchant}, P., {Fragos}, T., {Meynet}, G., \& {Kalogera}, V. 2019,
  \apjl, 870, L18, \dodoi{10.3847/2041-8213/aaf97b}

\bibitem[{{Ram{\'\i}rez-Agudelo} {et~al.}(2017){Ram{\'\i}rez-Agudelo}, {Sana},
  {de Koter}, {Tramper}, {Grin}, {Schneider}, {Langer}, {Puls}, {Markova},
  {Bestenlehner}, {Castro}, {Crowther}, {Evans}, {Garc{\'\i}a}, {Gr{\"a}fener},
  {Herrero}, {van Kempen}, {Lennon}, {Ma{\'\i}z Apell{\'a}niz}, {Najarro},
  {Sab{\'\i}n-Sanjuli{\'a}n}, {Sim{\'o}n-D{\'\i}az}, {Taylor}, \&
  {Vink}}]{ram17}
{Ram{\'\i}rez-Agudelo}, O.~H., {Sana}, H., {de Koter}, A., {et~al.} 2017, \aap,
  600, A81, \dodoi{10.1051/0004-6361/201628914}

\bibitem[{{Spera} \& {Mapelli}(2017)}]{spera2017}
{Spera}, M., \& {Mapelli}, M. 2017, \mnras, 470, 4739,
  \dodoi{10.1093/mnras/stx1576}

\bibitem[{{Spera} {et~al.}(2015){Spera}, {Mapelli}, \& {Bressan}}]{spera2015}
{Spera}, M., {Mapelli}, M., \& {Bressan}, A. 2015, \mnras, 451, 4086,
  \dodoi{10.1093/mnras/stv1161}

\bibitem[{{Spera} {et~al.}(2019){Spera}, {Mapelli}, {Giacobbo}, {Trani},
  {Bressan}, \& {Costa}}]{spera2019}
{Spera}, M., {Mapelli}, M., {Giacobbo}, N., {et~al.} 2019, \mnras, 485, 889,
  \dodoi{10.1093/mnras/stz359}

\bibitem[{{Spite} {et~al.}(2005){Spite}, {Cayrel}, {Plez}, {Hill}, {Spite},
  {Depagne}, {Fran{\c{c}}ois}, {Bonifacio}, {Barbuy}, {Beers}, {Andersen},
  {Molaro}, {Nordstr{\"o}m}, \& {Primas}}]{spiteetal05}
{Spite}, M., {Cayrel}, R., {Plez}, B., {et~al.} 2005, \aap, 430, 655,
  \dodoi{10.1051/0004-6361:20041274}

\bibitem[{{Spruit}(2002)}]{spruit2002}
{Spruit}, H.~C. 2002, \aap, 381, 923, \dodoi{10.1051/0004-6361:20011465}

\bibitem[{{Stevenson} {et~al.}(2019){Stevenson}, {Sampson}, {Powell},
  {Vigna-G{\'o}mez}, {Neijssel}, {Sz{\'e}csi}, \& {Mandel}}]{stevenson2019}
{Stevenson}, S., {Sampson}, M., {Powell}, J., {et~al.} 2019, arXiv e-prints,
  arXiv:1904.02821.
\newblock \doarXiv{1904.02821}

\bibitem[{{Stevenson} {et~al.}(2017){Stevenson}, {Vigna-G{\'o}mez}, {Mandel},
  {Barrett}, {Neijssel}, {Perkins}, \& {de Mink}}]{stevenson2017}
{Stevenson}, S., {Vigna-G{\'o}mez}, A., {Mandel}, I., {et~al.} 2017, Nature
  Communications, 8, 14906, \dodoi{10.1038/ncomms14906}

\bibitem[{{Sukhbold} {et~al.}(2016){Sukhbold}, {Ertl}, {Woosley}, {Brown}, \&
  {Janka}}]{sukhbold2016}
{Sukhbold}, T., {Ertl}, T., {Woosley}, S.~E., {Brown}, J.~M., \& {Janka}, H.-T.
  2016, \apj, 821, 38, \dodoi{10.3847/0004-637X/821/1/38}

\bibitem[{{Sukhbold} {et~al.}(2018){Sukhbold}, {Woosley}, \&
  {Heger}}]{sukhbold2018}
{Sukhbold}, T., {Woosley}, S.~E., \& {Heger}, A. 2018, \apj, 860, 93,
  \dodoi{10.3847/1538-4357/aac2da}

\bibitem[{{Takahashi} {et~al.}(2018){Takahashi}, {Yoshida}, \&
  {Umeda}}]{takahashi2018}
{Takahashi}, K., {Yoshida}, T., \& {Umeda}, H. 2018, \apj, 857, 111,
  \dodoi{10.3847/1538-4357/aab95f}

\bibitem[{{Tang} {et~al.}(2014){Tang}, {Bressan}, {Rosenfield}, {Slemer},
  {Marigo}, {Girardi}, \& {Bianchi}}]{tang2014}
{Tang}, J., {Bressan}, A., {Rosenfield}, P., {et~al.} 2014, \mnras, 445, 4287,
  \dodoi{10.1093/mnras/stu2029}

\bibitem[{{Uchida} {et~al.}(2019){Uchida}, {Shibata}, {Takahashi}, \&
  {Yoshida}}]{uchida2019}
{Uchida}, H., {Shibata}, M., {Takahashi}, K., \& {Yoshida}, T. 2019, \apj, 870,
  98, \dodoi{10.3847/1538-4357/aaf39e}

\bibitem[{{Ugliano} {et~al.}(2012){Ugliano}, {Janka}, {Marek}, \&
  {Arcones}}]{ugliano2012}
{Ugliano}, M., {Janka}, H.-T., {Marek}, A., \& {Arcones}, A. 2012, \apj, 757,
  69, \dodoi{10.1088/0004-637X/757/1/69}

\bibitem[{{van Loon} {et~al.}(2005){van Loon}, {Cioni}, {Zijlstra}, \&
  {Loup}}]{vanloonetal05}
{van Loon}, J.~T., {Cioni}, M. R.~L., {Zijlstra}, A.~A., \& {Loup}, C. 2005,
  \aap, 438, 273, \dodoi{10.1051/0004-6361:20042555}

\bibitem[{{Vartanyan} {et~al.}(2019){Vartanyan}, {Burrows}, \&
  {Radice}}]{vartanyan2019}
{Vartanyan}, D., {Burrows}, A., \& {Radice}, D. 2019, \mnras, 489, 2227,
  \dodoi{10.1093/mnras/stz2307}

\bibitem[{{Venumadhav} {et~al.}(2019){Venumadhav}, {Zackay}, {Roulet}, {Dai},
  \& {Zaldarriaga}}]{venumadhav2019}
{Venumadhav}, T., {Zackay}, B., {Roulet}, J., {Dai}, L., \& {Zaldarriaga}, M.
  2019, arXiv e-prints, arXiv:1904.07214.
\newblock \doarXiv{1904.07214}

\bibitem[{{Vink} {et~al.}(2000){Vink}, {de Koter}, \& {Lamers}}]{vink2000}
{Vink}, J.~S., {de Koter}, A., \& {Lamers}, H.~J.~G.~L.~M. 2000, \aap, 362,
  295.
\newblock \doarXiv{astro-ph/0008183}

\bibitem[{{Vink} {et~al.}(2001){Vink}, {de Koter}, \& {Lamers}}]{vink2001}
---. 2001, \aap, 369, 574, \dodoi{10.1051/0004-6361:20010127}

\bibitem[{{Vink} {et~al.}(2011){Vink}, {Muijres}, {Anthonisse}, {de Koter},
  {Gr{\"a}fener}, \& {Langer}}]{vink2011}
{Vink}, J.~S., {Muijres}, L.~E., {Anthonisse}, B., {et~al.} 2011, \aap, 531,
  A132, \dodoi{10.1051/0004-6361/201116614}

\bibitem[{{Woosley}(2017)}]{woosley2017}
{Woosley}, S.~E. 2017, \apj, 836, 244, \dodoi{10.3847/1538-4357/836/2/244}

\bibitem[{{Woosley}(2019)}]{woosley2019}
---. 2019, \apj, 878, 49, \dodoi{10.3847/1538-4357/ab1b41}

\bibitem[{{Woosley} {et~al.}(2007){Woosley}, {Blinnikov}, \&
  {Heger}}]{woosley2007}
{Woosley}, S.~E., {Blinnikov}, S., \& {Heger}, A. 2007, \nat, 450, 390,
  \dodoi{10.1038/nature06333}

\bibitem[{{Yoon} {et~al.}(2012){Yoon}, {Dierks}, \& {Langer}}]{yoon2012}
{Yoon}, S.~C., {Dierks}, A., \& {Langer}, N. 2012, \aap, 542, A113,
  \dodoi{10.1051/0004-6361/201117769}

\bibitem[{{Yusof} {et~al.}(2013){Yusof}, {Hirschi}, {Meynet}, {Crowther},
  {Ekstr{\"o}m}, {Frischknecht}, {Georgy}, {Abu Kassim}, \&
  {Schnurr}}]{yusof2013}
{Yusof}, N., {Hirschi}, R., {Meynet}, G., {et~al.} 2013, \mnras, 433, 1114,
  \dodoi{10.1093/mnras/stt794}

\bibitem[{{Zackay} {et~al.}(2019){Zackay}, {Venumadhav}, {Dai}, {Roulet}, \&
  {Zaldarriaga}}]{zackay2019}
{Zackay}, B., {Venumadhav}, T., {Dai}, L., {Roulet}, J., \& {Zaldarriaga}, M.
  2019, \prd, 100, 023007, \dodoi{10.1103/PhysRevD.100.023007}

\end{thebibliography}
\bibliographystyle{aasjournal}



\end{document}